Astronomy
&
Astrophysics

# Detection of new strongly variable brown dwarfs
# in the L/T transition⋆

Simon C. Eriksson, Markus Janson, and Per Calissendorff

Department of Astronomy, Stockholm University, 106 91 Stockholm, Sweden
e-mail: simon.eriksson@astro.su.se

**ABSTRACT**

*Context.* Brown dwarfs in the spectral range L9–T3.5, within the so called L/T transition, have been shown to be variable at higher amplitudes and with greater frequency than other field dwarfs. This strong variability allows for the probing of their atmospheric structure in 3D through multi-wavelength observations for studying the underlying physical mechanisms responsible for the variability. The few known strongly variable dwarfs in this range have been extensively studied. Now, more variables such as these need to be discovered and studied to better constrain atmospheric models. This is also critical to better understand giant exoplanets and to shed light on a number of possible correlations between brown dwarf characteristics and variability.
*Aims.* Previous studies suggest an occurrence rate for strong variability (peak-to-peak amplitudes >2%) of up to ~39% among brown dwarfs within the L/T transition. In this work, we aim to discover new strong variables in this spectral range by targeting ten previously unsurveyed brown dwarfs.
*Methods.* We used the NOTCam at the Nordic Optical Telescope to observe 11 targets, with spectral types ranging from L9.5 to T3.5, in the *J*-band during October 2017 and September 2018. Using differential aperture photometry, we then analysed the light curves for significant variability using Lomb-Scargle periodogram algorithms and least squares fitting.
*Results.* We report first discoveries of strong and significant variability in four out of the ten targets (false alarm probability <0.1%), measuring peak-to-peak amplitudes of up to $10.7 \pm 0.4\%$ in *J* for the T1 dwarf 2MASS J22153705+2110554, for which we observe significant light curve evolution between the 2017 and 2018 epochs. We also report a marginally significant detection of strong variability, and confirm that the well known 2MASS J01365662+0933473 is still strongly variable three years after the last reported epoch. Finally, we present an extensive multi-epoch catalogue of strong variables reported in the literature and discuss possible correlations that are identifiable from the catalogue.
*Conclusions.* We significantly add to the number of known strong variables, and through Poisson statistics infer an occurrence rate for strong variability among L9–T3.5 brown dwarfs of $40^{+32}_{-19}\%$, which is in agreement with previous estimates. The new variables identified in this work are also excellently suited for extensive multi-wavelength observations dedicated to probing the 3D structure of brown dwarf atmospheres.

infrared: stars – techniques: photometric

## 1. Introduction

Over the past two decades, research into brown dwarfs (BDs) has greatly expanded and diversified following an ever increasing number of discoveries of cool and ultra-cool dwarfs of spectral type L, T, and more recently Y. Studies of the all-sky near-infrared surveys such as The Two Micron All-Sky Survey (2MASS; Skrutskie et al. 2006), Wide-field Infrared Survey Explorer (WISE; Wright et al. 2010) and Sondage Infrarouge de Mouvement Propre (SIMP; Robert et al. 2016) have shown these objects to be quite ubiquitous. With the growing number of discoveries there were indications that as BDs cool and progress through late L and early T spectral types, they also undergo a rapid evolution in colour due to a brightening in the *J*-band with ~L8–T5 SpTs transitioning from positive *J − K* colour (red) towards negative (blue) with a shift of a few magnitudes, now commonly known as the L/T transition. This shift in colour is partly due to a suppression of flux in the *H*- and *K*-bands, caused by the onset of methane absorption in mid to late L-dwarfs (e.g. Noll et al. 2000; Helling & Casewell 2014), and due to *J*-band brightening likely caused by the occurrence of patchy clouds in the atmospheres of BDs (decreasing cloud opacity) as they transition from cloudy to clear photospheres (Ackerman & Marley 2001; Burgasser et al. 2002). Given the rapid rotation of BDs (e.g. Reiners & Basri 2008), this should give rise to a rotational modulation of the intrinsic brightness.

Early observations showed signs of photometric variability among L/T transition objects (Enoch et al. 2003), while the later high-precision photometry ($\sigma_{\rm phot} \sim 1\%$) studies starting with the T2.5 SIMP J013656.5+093347 (Artigau et al. 2009) enabled the observational testing of atmospheric models incorporating patchy clouds. It is interesting to note that J01365+0933 was shown to be periodically variable at ~2.4 h with peak-to-peak amplitudes of ~5%, making it the first BD where strong variability ($A > 2\%$) was robustly detected. This discovery was followed by the detection of the strongest variable to date, the T1.5 2MASS J21392676+0220226 (Radigan et al. 2012) showing peak-to-peak amplitudes up to 26% with periods of ~7.7 h. Results from both studies showed a wavelength dependence on the observed variability, with amplitude ratios $A_{K_S}/A_J < 1$, indicating that different wavelengths observe different layers in the atmosphere, offering a way to probe the 3D structure of brown dwarf atmospheres. Such multi-wavelength studies have since been shown to be increasingly important (Buenzli et al. 2012, 2014; Apai et al. 2013; Biller et al. 2013) and the key to determining which mechanisms give rise to variability. While larger ground surveys searching for variability (Girardin et al. 2013; Radigan et al. 2014; Vos et al. 2019) are effective at discovering and partly characterising new variables, the use of simultaneous *Hubble* Space Telescope (HST) and *Spitzer* Space Telescope (SST) observations offer a unique opportunity to effectively probe the atmospheric 3D structure in depth with very high precision and long baselines (Yang et al. 2016; Apai et al. 2017; Biller et al. 2018).

While the study of giant exoplanet atmospheres requires high-contrast imaging using large ground based telescopes, field dwarfs on the other hand are generally easier to observe. Given similarities in temperature, surface gravity and colour between young giant exoplanets and young BDs, the latter has been suggested as easily observed analogues for studying the atmospheres of the former (Biller et al. 2015; Vos et al. 2019). Furthermore, the variability mechanisms is not believed to be different for young dwarfs compared to older field dwarfs, indicating that the study of BD atmospheres in general could be of significant value to understanding the atmospheres of the much harder to image giant exoplanets. Finally, the very recent and interesting multi-wavelength study of variability in Jupiter by Ge et al. (2019) indicates that the mechanisms driving variability in Jupiter likely share similarities with those in BDs.

While the 44 L3–T8 BD survey by Metchev et al. (2015) infers that nearly all L- and T-dwarfs should be variable with amplitudes at ≥0.2%, the survey of 62 L4–T9 BDs by Radigan et al. (2014) showed that strong variables are significantly more common within the L/T transition, inferring that ~39% of L9–T3.5 BDs should be strongly variable. As peak-to-peak amplitudes are known to vary (significantly) with wavelength, strong variables offer the best chances for effectively probing the 3D structure of atmospheres. The half a dozen strongest L/T transition variables currently known have all been observed in detail (Yang et al. 2016; Apai et al. 2017), showing a great variety in short- and long-term light curve evolution leading to new hypotheses about the underlying causes. Light curves too complex to be modelled by a single rotating spot are better explained by a combination of light cloud bands modulated by planetary-scale waves and spots (Apai et al. 2017), similar to what has been observed in a very recent multi-wavelength study of Jupiter (Ge et al. 2019), indicating that aspects of the underlying mechanisms could shared with BDs. A number of trends and possible correlations have also been identified among the currently known variables. Several strong variables, including J2139+0220, have been previously (mis-) identified as unresolved binaries, possibly indicating that the index criteria used for detecting unresolved binarity could be useful for identifying likely variable candidates (Manjavacas et al. 2019a). Phase shifts between the light curves in different wavelengths have been observed (e.g. Buenzli et al. 2012; Yang et al. 2016), and there are also significant correlations between variability amplitude and various properties such as inclination and colour (Vos et al. 2017). To advance our understanding of these mechanisms and possible correlations, more strongly variable brown dwarfs need to be discovered and studied in depth.

## 2. Observations and data reduction

In this section we discuss our target selection and present target information and the accompanying observation log from two epochs at the NOT. As this is the first time the NOT has been used for this type of variability study, we also describe the reduction process in depth, concluding with aperture photometry details.

### 2.1. Target selection

Based on the findings by Radigan et al. (2014) that strong variability mainly occurs in the L9–T3.5 SpT range within the L/T transition, and the need for new variable T-dwarfs to study in depth as done by Yang et al. (2016) and Apai et al. (2017), we focus our study entirely on this SpT range. If patchy clouds are primarily responsible for variability, then this range should be ideal as it is during the L/T transition that cloud layers are disrupted, revealing lower, hotter regions of the atmosphere (Ackerman & Marley 2001; Burgasser et al. 2002). This focus allows us to directly compare our results with the survey by Radigan et al. (2014), the largest ground-based variability survey to date, also performed with a telescope of the same size as the one used in our survey. The majority of the targets are selected from the SIMP brown dwarf survey of 169 M-, L- and T-dwarfs by Robert et al. (2016), where all L/T transition dwarfs observable from Observatorio del Roque de los Muchachos on La Palma during the autumn months are included in Table 1.

J0013−1143, a T3 dwarf recently discovered by Kellogg et al. (2017), was added to the list for our second epoch observations. The authors classify J0013 as a likely unresolved binary with a T3.5 + T4.5 composite spectra, as it better matches the spectral fit and satisfies two of the six binary index criteria described by Burgasser et al. (2010a). They do however consider that it could instead be a variable T-dwarf displaying two temperature components. As is discussed in Sect. 5, there seems to be a trend of some BDs previously classified as unresolved binaries, using such index criteria selection, turning out to be strong variables, see for example J1324+6358 (Metchev et al. 2015; Yang et al. 2016) or J2139+0220 (Radigan et al. 2012) from Table D.1 in Sect. 5.1. Our findings for J2215+2110 (T1, classified as an unresolved binary by Kellogg et al. 2015) from the first epoch seemed to support this trend, and with the addition of J0013 the study now included three targets with composite atmosphere spectral features.

### 2.2. Nordic Optical Telescope NOTCam

The observations listed in the observation log (Table 2) took place between 2017 and 2018 with the NOTCam instrument at the 2.5 m Nordic Optical Telescope (NOT), located at Observatorio del Roque de los Muchachos on La Palma. All observations were run in visitor mode, except for J0136+0933 which ran in service mode as compensation for time lost to target of opportunity (ToO) interruption. The five-night observing run in October 2017 was heavily impacted by poor weather conditions, yielding a total of nine hours of usable data. The following four-night run in September 2018 had substantially better conditions with 28 h of effective on-sky time, with only sporadic clouds during the final two nights.

NOTCam was used with the $J$ filter ($1.165-1.328\,\mu$m) in wide field imaging mode, where the $1024 \times 1024$ detector has a pixel scale of $0.234''$ resulting in a $4' \times 4'$ field of view. The

**Table 1.** Identifier, SpT and colour information for targets observed in this work.

| Target 2MASS ID | Compact ID | SpT [a] (NIR) | Ref. [b] | 2MASS | | | $d_{phot}$ [c] (pc) |
|---|---|---|---|---|---|---|---|
| | | | | $J$ (mag) | $K_S$ (mag) | $J - K_S$ (mag) | |
| 2MASS J00132229−1143006 | J0013−1143 | T3 (T3.5+T4.5?) | 4, 4 | $16.35 \pm 0.10$ | $15.98 \pm 0.22$ | 0.37 | ... |
| 2MASS J01352531+0205232 | J0135+0205 | L9.5 | 3, 3 | $16.62 \pm 0.13$ | $15.12 \pm 0.12$ | 1.50 | $29.6 \pm 4.8$ |
| 2MASS J01365662+0933473 | J0136+0933 | T2.5 | 2, 7 | $13.45 \pm 0.03$ | $12.56 \pm 0.02$ | 0.89 | $5.9 \pm 1.0$ |
| 2MASS J01383648−0322181 | J0138−0322 | T3 | 6, 6 | $15.30 \pm 0.08$ | 1.06 | $18.0 \pm 2.9$ |
| 2MASS J01500997+3827259 | J0150+3827 | L9.5 | 7, 7 | $16.11 \pm 0.08$ | $14.48 \pm 0.07$ | 1.63 | $19.4 \pm 3.1$ |
| 2MASS J03162759+2650277 | J0316+2650 | T2.5 (T2+T7.5?) | 3, 3 | $16.11 \pm 0.02$ [d] | $15.50 \pm 0.05$ [d] | 0.61 [d] | $23.1 \pm 3.8$ |
| 2MASS J21324898−1452544 | J2132−1452 | T3.5 | 1, 7 | $15.71 \pm 0.09$ | $15.27 \pm 0.18$ | 0.44 | $18.2 \pm 2.9$ |
| 2MASS J21483578+2239427 | J2148+2239 | T1 | 7, 7 | $16.46 \pm 0.12$ | >14.90 | <1.56 | $26.1 \pm 5.0$ |
| 2MASS J22153705+2110554 | J2215+2110 | T1 (T0+T2?) | 5, 5 | $16.00 \pm 0.08$ | $14.82 \pm 0.11$ | 1.18 | $20.9 \pm 3.4$ |
| 2MASS J22393718+1617127 | J2239+1617 | T3 | 6, 7 | $16.08 \pm 0.08$ | >14.89 | <1.19 | $17.8 \pm 2.9$ |
| 2MASS J23032925+3150210 | J2303+3150 | T3 | 8, 7 | $16.22 \pm 0.09$ | $15.44 \pm 0.16$ | 0.78 | $23.5 \pm 3.8$ |

**Notes.** For this work we have chosen to use 2MASS identifiers whenever possible and alternate designations (from e.g. the PSO, SIMP or WISE surveys) might be used in discovery references. Compact ID entries in the table are of the form Jhh:mm±dd:mm and in running text also frequently referred to as Jhh:mm. [a] Possible binarity indicated in parenthesis with question mark. [b] References: discovery, spectral type. [c] Photometric distances from Robert et al. (2016). [d] MKO magnitudes from Best et al. (2015) instead of 2MASS.
**References.** (1) Andrei et al. (2011); (2) Artigau et al. (2006); (3) Best et al. (2015); (4) Kellogg et al. (2017); (5) Kellogg et al. (2015); (6) Kirkpatrick et al. (2011); (7) Robert et al. (2016); (8) Schneider et al. (2016a).

**Table 2.** Observation log observations using NOT/NOTCam $J$.

| Target | Date (UT) | $\Delta t$ (h) | $t_{exp}$ (s) | $SD_{raw}$ [a] |
|---|---|---|---|---|
| J0013−1143 | 2018-09-28 | 2.95 | 150 | 0.14 |
| J0135+0205 | 2018-09-28 | 2.44 | 150 | 0.14 |
| J0136+0933 | 2018-07-30 | 2.76 | 60 | 0.05 |
| J0138−0322 | 2017-10-11 | 1.55 | 150 | 0.24 |
| | 2018-09-25 | 4.07 | 120 | 0.11 |
| J0150+3827 | 2018-09-26 | 1.72 | 160 | 0.04 |
| J0316+2650 | 2018-09-26 | 2.42 | 150 | 0.04 |
| J2132−1452 | 2017-10-11 | 4.12 | 150 | 0.20 |
| | 2018-09-27 | 1.11 | 180 | 0.11 |
| J2148+2239 | 2017-10-12 | 0.84 | 150 | 0.25 |
| | 2018-09-26 | 4.19 | 150 | 0.06 |
| J2215+2110 | 2017-10-13 | 2.40 | 240 | 0.37 |
| | 2018-09-25 | 5.15 | 90 | 0.06 |
| J2239+1617 | 2018-09-26 | 2.64 | 150 | 0.21 |
| J2303+3150 | 2018-09-28 | 1.64 | 150 | 0.50 |

**Notes.** Targets observed during NOT programs 56-002 and 57-006, listed in descending (compact) 2MASS ID. For more target information see Table 1. The date represents the start of the night where the observation was executed. [a] The standard deviation of the normalized raw target light curve obtained from aperture photometry. Shown here as an indicator of the quality of the observing conditions, with detailed plots of the normalised reference light curve, sky counts and airmass being available in Appendix A. For observations where $t_{exp}$ was changed shortly after the beginning of the run, the listed SD value is based on the frames using the final $t_{exp}$.

$J$-band was chosen as it offers a favourable wavelength range for a combination of high target brightness and high expected variability amplitude. Our targets are generally brighter in the $K$-band, as is the thermal background, but numerous studies indicate a lower expected amplitude in $K$ ($A_K / A_J \lesssim 0.5$, e.g. Artigau et al. 2009; Radigan et al. 2012; Vos et al. 2019).

Two to three targets per night were observed in a nine point dither ($3 \times 3$ grid roughly 18″ in diameter) around the starting position, with one exposure at each location corresponding to a total integration time of NDIT×DIT (e.g. $6 \times 10$ s for J0136+0933). NDIT and DIT were varied as needed to keep the target and sky counts below 22 000 Analog-to-Digital Units (ADUs) to minimise non-linearity effects, read noise and to keep the total exposure time low enough for effective sky subtraction or to avoid over-saturating nearby bright stars in the field.

Typical seeing at the NOT is 0.6–1.5″, but as real-time seeing information is not recorded in the file headers, we instead present in Table 2 the standard deviation of the normalised raw target light curve at the chosen aperture size, $SD_{raw}$, as an indicator of the quality of the observing conditions. Using Python[1] we also approximate the seeing during our observations by measuring the FWHM of the stellar point spread functions (PSFs) of stars in the field, yielding an average FWHM of 0.7″. This information is later used to verify that there are no significant correlations between changes to for example FWHM or airmass and the target or reference star light curves after differential photometry (Sect. 3.1) has been performed. Other indicators such as the normalised reference light curve, sky-subtracted background flux and airmass information are available for all targets and epochs in Appendix A. Finding charts are available in Appendix C.

The reduction of NOT/NOTCam data involves steps common to most near-infrared (NIR) image reduction and we outline these and some considerations specific to the NOTCam array below. Steps Sects. 2.2.1–2.2.3 were executed automatically in a pipeline written in Python 3.7, followed by standard aperture photometry on the reduced frames.

### 2.2.1. Non-linearity correction

Among our 11 targets, only J0136+0933 was at risk of exceeding 22 000 ADUs due to its brightness and lack of equally bright stars in the field. The remainder of the targets typically fell into the 3000–10 000 ADU range where the non-linearity of the detector is on average between 0.05 and 0.23%. Since the detector is to some extent non-linear at all counts, and some bright reference stars exceed 22 000 ADUs we implement a non-linearity correction as described by the NOTCam user pages at not.iac.es. The correction is applied pixel by pixel to the raw image using two coefficient images $b_a$ and $c_a$ and the polynomial $y = x(1 + b_a x + c_a x^2)$. Applying the correction does not give rise to any known negative effects and works well up to at least 46 000 ADUs. Given the reported reliability of the correction, reference stars with counts up to this level, which is also comfortably below the 56 000 ADU saturation level, were allowed in the differential photometry process.

---

[1] IRAFStarFinder in the package phototutils.

### 2.2.2. Flat-fielding

Dithered twilight skyflats were used to create a master differential flatfield image from bright and faint images, subtracting out the dark. The faint image is subtracted from the bright image, with both taken at the same pointing within less than half an hour of each other. If conditions and the resulting flats were poor, a master flat from another night closest in time was used instead.

### 2.2.3. Sky subtraction

For imaging in NIR bands like $J$, a good sky subtraction is often critical to maintaining photometric precision over a longer observation, as the high background sky counts often exceed those intrinsic to the target. As most of our targets are relatively faint, the desire for a high ADU count is weighed against the need for efficient sky subtraction. If conditions are also poor, finding an optimal balance between the two is sometimes difficult. For a given nine-point dither sequence, taking the median of a given pixel among all frames in the series is the optimal method to create a median combined sky frame, but only if the time between each frame is sufficiently short. As a result of some of our targets requiring longer total integration times to reach acceptable ADU counts, sky subtraction was done using either two (for $t_{exp} > 120$ s) or four of the dithering points closest in time.

The result was an overall stable sky subtraction, where over-subtraction sometimes excluded reference stars in crowded fields. Depending on observing conditions, frames were sometimes discarded due to excessive cloudiness or background counts, as seen for the J2239+1617 or J0136+0933 observations in Appendix A).

### 2.2.4. Bad pixels and detector defects

The NOTCam array has two columns of zero-value pixels at $x = 512, 1024$, of which the middle one inevitably leads to the loss of some potential reference stars than end up on it. For one epoch of observing J2132−1452, one frame from each nine-point dither was lost to this issue.

In addition to these columns, the array has a number of other zero-value pixels at the edges, as well as isolated or groups of cold pixels with a response of 20–40% of unaffected pixels. Should a target or reference star end up in an area of the array more commonly populated by these, the dithering motion can give rise to a seemingly periodic variability. This variability will however be of low amplitude and always display a period equal to the time it takes to perform a full nine-point dither with overheads, up to 40 min for these observations. This period is much shorter than the period of the fastest rotating BD with detected variability (1.4 h, J2228−4310; Clarke et al. 2008; Radigan et al. 2014), can be easily identified with periodogram analysis and is effectively corrected for using differential photometry (see Sect. 3.1). Correction of bad pixels by interpolation was ruled out as we wanted to be able to clearly see which objects were affected, and thus unreliable, and did not want to risk altering the photometry through such action.

The array experiences an optical distortion effect which radially elongates the PSF as a function of distance from the edge of the array. Since only objects close to the edge are significantly affected this has a limited effect on our observations. The optical distortion correction described by the NOTCam user pages detailing calibration[2] was tested on a number of our observations (e.g. J0136 and both epochs of J2215). The distortion effect was corrected with the IRAF geometric transformation routine geo-tran run in PyRAF, using the spatial transformation database provided by the NOT[3]. The outcome was only marginally better or worse photometry. Therefore, with no clear benefits to applying the procedure, and to avoid unnecessarily altering data, we opted not to use the correction for the final analysis.

### 2.2.5. Aperture photometry

Following the above data reduction the dithered frames were re-centred prior to aperture photometry, done by full-pixel shifts (conserving flux) to a common centre typically defined from the first dither position. Next, fixed aperture photometry using inner aperture radii of three to seven pixels ($\sim 1-2.3$ times the typical measured stellar PSF FWHM) in 0.5 pixel increments, was run through two independent routines: photutils in Python and apphot in IRAF. The outer sky aperture was defined by an annulus with an inner radius of ten pixels and an outer of 20, containing $\sim 800$ pixels for the purpose of residual background subtraction. Centroiding the apertures is done by apphot.center in IRAF and DAOStarFinder in photutils. In general, a larger inner aperture is desirable for brighter targets, but for fainter targets the benefit of collecting photons from the wings of the PSF is outweighed by a greater contribution from background noise. For consistency, the same aperture size is used for both the target and reference stars for a given field and epoch. The final aperture size is arrived at iteratively through the differential photometry process (Sect. 3.1). This aperture size is also listed for a given epoch in the plots detailing local observing conditions in Appendix A.

The initial selection process of reference stars needed for the differential photometry is done outside of the reduction pipeline with manual input. (I) Any source with an extended PSF, such as a galaxy or merged stellar PSFs, are rejected. (II) Due to the sky subtraction process, stars in close proximity on the array might be affected by over-subtraction from a neighbour during dithering and any such stars are rejected. (III) Stars that suffer from saturation or are above the 46 000 ADU limit for the non-linearity correction are rejected. For data processing and reference star numbering, the target is set as the first object, with the reference stars following in decreasing brightness, numbered 1–n (see finding charts in Appendix C).

Both photutils and apphot provided photometry of equivalent quality for bright targets such as J0136+0933 or observations with excellent conditions, for example J2215+2210 in 2018. However, photutils struggled to produce reliable results for fainter targets or during poor conditions, so the final analysis was done using photometry and photometric uncertainties from apphot in PyRAF.

---

[2] http://www.not.iac.es/instruments/notcam/calibration.html
[3] http://www.not.iac.es/instruments/notcam/distortion/

## 3. Light curve analysis

The raw light curves obtained from aperture photometry are affected by brightness fluctuations greater than the amplitude one might expect from a strongly variable BD and must be corrected prior to any meaningful analysis. For this purpose we follow the basic idea behind relative differential photometry, as described by Radigan et al. (2012). These fluctuations arise from alterations made to $t_{exp}$, instrumental effects, changing atmospheric opacity, for example due to clouds and changes in seeing or airmass. As all objects in the field are similarly affected, these effects can be corrected for to a level similar to the photometric precision (∼1%) by using differential photometry, where the raw target light curve is divided by a calibration curve from a set of stable reference stars from the field. During the process, the calibrated relative flux light curve of the target is analysed for significant variability using two Lomb-Scargle periodogram (Lomb 1976; Scargle 1982) routines and fit using a Levenberg-Marquardt least-squares algorithm (Levenberg 1944; Marquardt 1962). This process is detailed below and the summarised results can be viewed in Table 3 (Sect. 4).

### 3.1. Differential photometry

Once reference stars have been selected and aperture photometry performed as outlined in Sect. 2.2.5, the relative differential photometry process is performed in Python in combination with the full light curve analysis. The process is executed as follows. (I): the light curves of up to eight reference stars (RS) are included and each light curve is normalised by its own median flux. (II): or each RS, a calibration curve is obtained by median combining the light curves of all other included RS, so that an individual RS or the target BD is excluded from its own calibration. The light curves of the other RS are stacked in an array and the median flux in each frame is extracted to form the calibration curve which the individual RS light curve is then divided by. These are then the relative flux light curves for the target and reference stars shown in for example the bottom panels in Fig. 3. (III): Once differential photometry has been performed, the target light curve is automatically fit with a Levenberg-Marquardt (LM) algorithm for a linear and sinusoidal model. The target and RS light curves are also analysed by a Generalised Lomb-Scargle Periodogram (GLSP) and a Bayesian version (BGLSP) as described in Sect. 3.2 below. (IV): Here the process becomes iterative. Poor quality RS, for example those that show a high standard deviation (SD) of the light curve due to intrinsic variability (indicated by eye or GLSP) or high scatter from poor photometric precision ($\sigma_{phot} \gtrsim 2\%$), are excluded and if $N_{RS} \leq 8$, steps (I)–(III) are iterated until a final stable set of three to six RS is obtained. For $N_{RS} > 8$, (I)–(III) is first iterated until the most stable eight RS are identified, after which the set is further refined down to the best three to six RS by removing lower quality RS.

The final set(s) should ideally also be the most stable for different aperture sizes, as fainter and brighter objects typically have greater scatter at larger (∼5–6 pixels) and smaller (∼3–4 pixels) apertures respectively. This trend tends to be reinforced during less than optimal observing conditions. Once a set of suitable references has been identified, the iterations continue in an effort to identify the optimal aperture size. An initial guess is obtained from $\sigma_{phot}$ (photometric precision estimated by apphot) which, especially for fainter targets, will give a good indication at which aperture size(s) the photometry is most stable. To estimate the scatter of the target light curve even in the presence of significant variability, the mean of the SDs of the light curve divided into eight bins is used. This mean SD should converge towards the SD of the residual (data – best fit) for an optimal linear or sinusoidal fit.

With a good field of reference stars, it is often the case that the most suitable final set is composed of stable stars of similar brightness to the target. In this case the process will generally conclude with an aperture size that is ±0.5 pixels from the initial guess based on $\sigma_{phot}$, as the photometric precision tends to be the main contributing factor for scatter in the data given that our targets are rather faint. In other cases there might not be sufficiently many, if any, stable stars of similar brightness. In that event an optimal compromise in photometric precision between the target and reference stars needs to be reached, especially in the presence of very bright RS which can be unstable at smaller apertures. This can therefore require several more iterations for the purpose of arriving at an aperture size that offers an acceptable level of photometric precision for the target and stability in the reference stars.

Figure 1 illustrates differential photometry in practice, going from the raw flux to the calibrated relative flux, for three observations with varying observing conditions. J0136+0933 was observed during excellent conditions (Appendix A) and the trend of increasing flux is primarily due to decreasing airmass. J2215+2210 (2017) and J2239+1617 (2018) were observed during partly cloudy conditions and the plots show the efficiency of the method for removing global effects.

We choose to do the full analysis even for sets that include lower quality reference stars, rather than first identifying the most stable based on their SD and $\sigma_{phot}$ and then continuing with fitting and periodogram analysis. Doing so aids in identifying trends from for instance dithering effects but also allows the analysis to more confidently converge on the true shape of the target light curve. Even a set of poor quality RS used in differential photometry will still effectively remove the fluctuations in the raw target flux. While they might also induce trends or variability not intrinsic to the target, the target light curve obtained from using RS with (very) poor photometric precision will generally be similar to that obtained from a set of high quality RS. We exemplify this in Sect. 4.1 during the analysis of J0138, which lacks high quality reference stars in the field, by showing that the variability in J0136 is significant even if only two poor quality RS are used.

### 3.2. Detecting significant variability

As described above, even poor reference stars work well for removing the brightness fluctuations of the raw light curve. The challenge lies in arriving at an optimal combination of a set of RS and an aperture size that allows one to confidently claim whether a target is significantly variable or not for a given epoch. Significant detections can provide us with accessible "testing grounds" for atmospheric physics models, but as more and more BDs are surveyed, reliable non-detections also become increasingly important. As we discuss in Sect. 5.1, there are many characteristics thought to be more commonly associated with variable BDs than non-variable BDs, but to be able to verify these possible correlations, confident non-detections are also required.

An efficient first step in analysing a light curve for periodic (sinusoidal) signals is through the use of Lomb-Scargle periodograms (VanderPlas 2018). There exists a number of different algorithms and for this work we have chosen to include two of them in the analysis, the Generalised Lomb-Scargle periodogram (GLSP) and the Bayesian Generalised Lomb-Scargle periodogram (BGLSP) as described by Zechmeister & Kürster (2009) and Mortier et al. (2015) respectively. The GLSP improves on the regular LSP by using a floating mean and weights from observation uncertainties when fitting the data. It was pointed out by Manjavacas et al. (2017) that observational gaps in a time series could result in incorrect periods being favoured by the regular GLSP, a consideration also taken into account in Vos et al. (2019). Since the observations of J2239+1617 (2018) includes a large gap due to clouds and J2132−1462 (2017) contains several smaller gaps due to bad frames we apply the BGLSP as suggested in Manjavacas et al. (2017), as a precaution to ensure we obtain the true periods for a given data set and achieve parity in robustness with previous and upcoming surveys. The Python package astropy.stats includes a LSP routine, which when run with the method = "slow" option and photometric uncertainties as weights applies the GLSP described above. Mortier et al. (2015) made their BGLSP routine in Python publicly available which we have incorporated into the analysis code. Both periodograms were run with minimum periods equal to a full dither sequence without overheads ($9t_{exp}$) and maximum periods equal to twice the observation length ($2\Delta t$).

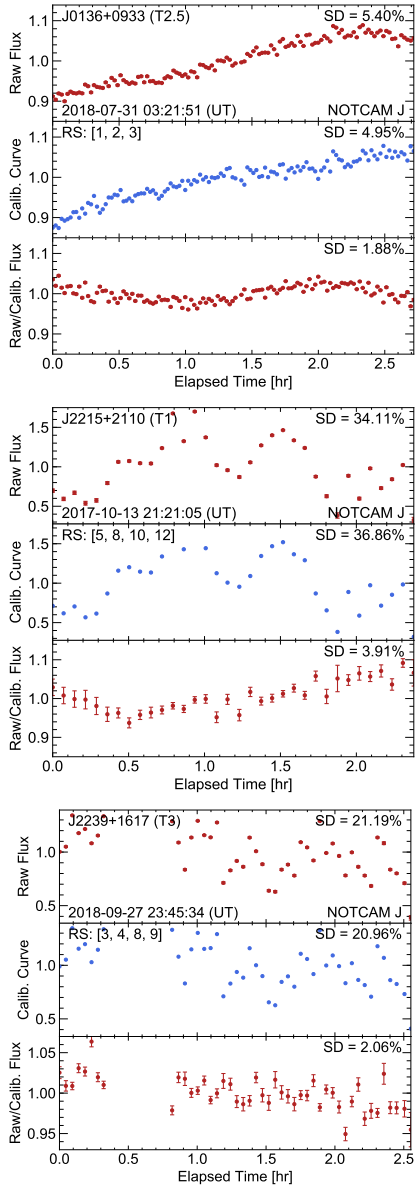

**Fig. 1.** Three-panel figures showing the differential photometry at work for J0136, J2215 and J2239 respectively. Axis scales are preserved between panels for illustrating the efficiency of the process, and detailed plots of detected variability are available in Sect. 4. (I) *Top panel*, in a given figure: normalised raw flux of the target after aperture photometry, with error bars and the standard deviation (SD) of the light curve. (II) *Middle panel*: calibration or reference light curve created by median combining the normalised raw light curves of the reference stars (RS) indicated in the panel. (III) *Bottom panel*: relative differential photometry of the target, obtained by dividing the top panel light curve with the middle panel light curve.

The process is very efficient in detecting possible periods, but does not validate the statistical significance of these detections. For Lomb-Scargle periodograms this is achieved by calculating a false alarm probability (FAP) at a given level of significance. There are a number of methods for computing the FAP and from these we apply the bootstrap method, considered the most robust albeit also the most computationally expensive (VanderPlas 2018). Calculating the FAP with this technique involves randomly resampling the data $N_{boot} \approx 10/P_{FAP}$ times for a given periodogram, which for a FAP of 1% (99% significance) means 1000 resamplings. This is similar to the method applied by Radigan et al. (2014) where they define a 99% significance from 1000 simulated light curves, obtained by randomly permuting indices of the originals, and is also applied by Vos et al. (2019). However, both Radigan et al. (2014) and Vos et al. (2019) find that >1% of their simulated light curves fall above the 1% FAP level, the cause of which they attribute to residual noise, and adjust their >99% significance to 1.4–3.7 times the 1% FAP depending on the instrument used for observations. Taking careful note of these results, we define our >99% significance level at a FAP of 0.1% ($N_{boot} \approx 10^4$) and calculate it using Python[4]. For a given observation, the GLSP is computed for the target, selected reference stars and the (data-fit) residual for the target. The GLSP of the residual gives an indication of how well the period of the variability was isolated by the LM fit, and both the 1% and 0.1% levels are indicated by dashed horizontal lines in GLSP presented in this work. The BGLSP is used to verify that the probable period indicated by the GLSP is the true one for a given observation. For targets with sinusoidal variability, a Gaussian can be fit to very strong peaks in the BGLSP to estimate the uncertainty from the FWHM at the peak period, which can be compared to the period obtained from the LM fit. We illustrate the result of this procedure in Fig. 2.

The Python routine "lmfit" was used to fit a linear ($y = a + bx$) and sinusoidal ($F(t) = a + A\sin(t\omega + c)$) function to the target light curve and estimate the peak-to-peak amplitude. From a by-eye estimate alone it is often clear if the variability is sinusoid-like or a linear trend but for ambiguous cases and for optimising RS selection it is useful to calculate the SD of the residual ($SD_{res}$) and compare it with the SD of the light curve ($SD_{data}$). A further SD statistic is calculated as the mean of the SD of the light curve divided into eight bins ($SD_{bin}$), as this is representative of the scatter in the data even in the presence of strong variability. The best fit will maximise the quantity $SD_{data} - SD_{res}$, with $SD_{res}$ ideally approaching the value of $SD_{bin}$. References with a photometric precision worse than the target will typically increase scatter in the light curve and therefore also $SD_{bin}$, as is the case for J0136+0933 where $SD_{bin} \approx 10 \times \sigma_{phot}$ due the target being significantly brighter than the available references.

---

[4] FAP calculated with the routine false_alarm_level in astropy.stats/ LombScargle using method = "bootstrap".

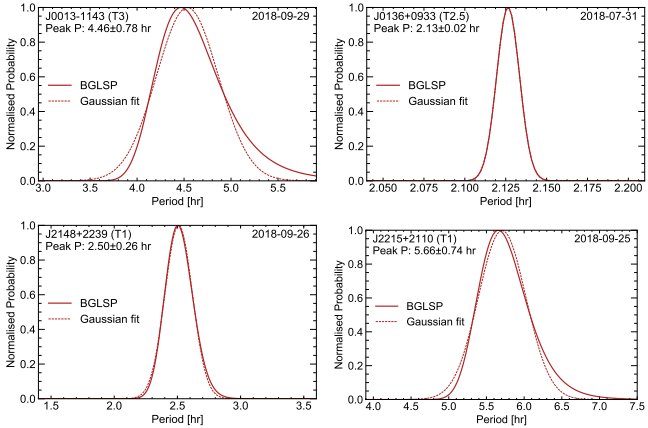

**Fig. 2.** Examples of four Bayesian generalised Lomb-Scargle periodograms (BGLSP) from the significant detections of variability in J0013, J0136 and J2215, as well as the marginally significant detection in J2148, showing normalised probability and the Gaussian fits used to estimate the rotation period $1\sigma$ uncertainty by calculating the FWHM.

For an optimal sinusoidal fit, reported uncertainties by lmfit in $a$, $\omega$ (and thus the rotation period) and $c$ should be minimised. Combined with the analysis of the periodograms where the target should share as few frequencies as possible with the references, the result is a robust method for detecting significant variability in brown dwarfs.

## 4. Results

We detect significant (>99%) variability in five out of the 11 targets with peak-to-peak amplitudes of at least 2%, with four out of these five representing new discoveries of strongly variable T-dwarfs in the L/T transition. We detail these discoveries below, present key light curves and periodograms in Figs. 3 and 5, summarise all results in Table 3 and put them into a wider context in the Colour vs SpT diagrams in Figs. 8 and 7. One additional target displays clear signs of strong variability, but falls just short of our statistical threshold to qualify as significant with a FAP of 1.9% (Fig. 6). The remaining five targets showed either no signs of variability (3/5) or were tentatively variable with high FAP values (2/5). Plots detailing observing conditions are available in Appendix A and relevant plots for non-significant detections are available in Appendix B.

### 4.1. Significant detections of variability

We first present our analysis of the five significant detections of strong variability, divided into subsections by ascending 2MASS ID.

#### 4.1.1. 2MASS J00132229−1143006

As mentioned in Sect. 2.1, this blue ($J - K_S = 0.37$) T3 dwarf was flagged by Kellogg et al. (2017) as a likely unresolved binary (T3.5 + T4.5), making it an interesting target for this survey. We observed J0013−1143 in visitor mode during the night of 2018-09-28 under good conditions (Appendix A) for almost three hours. The initial drop in flux is due to a decrease in $t_{exp}$ with other fluctuations mainly being due to slowly increasing airmass and a few sporadic thin clouds.

For this target we detect strong ($A \geq 2\%$) and highly significant variability (FAP ≪ 0.1%), indicated by the GLSP and light curves shown in Fig. 3. Out of the available ten reference stars we select four for the final light target curve, which shows a similar trend as the initial iterations using sets of eight RS, but with less RS influence in the GLSP. LM fitting favours a linear trend with an amplitude of 4.6±0.2% for the observation. A sinusoidal fit is also possible but is more unstable across aperture and RS selection, and yields a highly uncertain period with a lower reduction of the residual SD than the linear fit. Both the GLSP and BGLSP (top left in Fig. 2) indicate a peak-power period of $4.5 \pm 0.8$ h, but restricting the LM algorithm to intervals around this period does not produce a viable fit for any RS selection. Given the apparent linear trend, which eliminates virtually all power in the GLSP when removed, and instability of attempted sinusoidal fits we define a minimum period of 2.8 h where the GLSP crosses the 0.1% FAP, but recognise that the true rotation period is likely at least four to five hours, assuming a single peaked light curve. As we don't seem to observe neither a peak nor trough, the amplitude measurement represents a minimum value, but the implication that J0013 is a (very) strong variable while also having been flagged for unresolved binary gives further indication that there might be a correlation between the two (Sect. 5).

#### 4.1.2. 2MASS J01365662+0933473

First discovered by Artigau et al. (2006) as SIMP J013656.57+ 093347.3 and later found to be strongly variable by Artigau et al. (2009), the T2.5 BD J0136+0933 was observed in service mode as ToO compensation during the night of 2018-07-30 for 2.8 h, primarily to verify its continued variability but also to serve as a control for the validity of our analysis. Conditions were excellent with flux increasing during the observation due to decreasing airmass.

We find that J0136 continues to be a strong variable (middle in Fig. 3) with high significance. From eight available references the three brightest stable (but still not of equal brightness to the target) stars were chosen for the final set. A lack of equal brightness RS in the field gives rise to a scatter in the data much greater than that expected from photometric precision alone, but which is still random in nature and thus does not greatly affect the analysis. The trend is clearly sinusoidal, not present in any RS and is easily fit for with a peak-to-peak amplitude of 4.4±0.2% and a peak-power period of 2.13 h in the GLSP, which is efficiently subtracted through the fit. The BGLSP (top right in Fig. 2) indicates a very narrow distribution of periods centred on $2.13 \pm 0.02$ h, compared to $2.14 \pm 0.05$ h obtained from the LM fit. This close agreement in period measurements to within 0.5% demonstrates the reliability and robustness of an analysis based on the combination of these three methods.

However, as can be seen from our cataloguing of strong variables in Table D.1, this period deviates by over 10% from several previous observations of J0136 that indicate a period of ∼2.4 h. The likely reason for this discrepancy might be found in the works by Artigau et al. (2009) and in the supplementary material of Apai et al. (2017). Artigau et al. (2009) observe a "split" peak in J0136 during their September 18, 19 and 21 observations. For the 6th *Spitzer* visit in Apai et al. (2017) the light curve of J0136 is similarly double peaked over several rotations, with an additional peak shortly after the highest one. In the same material extensive light curves of J1324+6358 (T2) and J2139+0220 (T1.5) also display a merger and splitting of the main peak. As can be seen in the light curves (Artigau et al. 2009), the split peaks also drift apart slightly from rotation to rotation, being closest on September 18, furthest apart in September 19 and then closer on September 21. The cause for this drift could be the planetary-scale waves described in Apai et al. (2017); for our observation it seems possible that we start our run at this second lower peak and therefore do not sample a full rotation. The seeming lack of a second peak at the end could mirror the case for the September 18 observation in Artigau et al. (2009), where the first rotation seen during this epoch shows a single peak, while the next indicates the start of a split. There are a further two epochs of J0136 in Croll et al. (2016) where a narrowly split peak is visible, and others with wider splits. For our epoch of J0136 we choose to adopt the period and amplitude obtained from the LM fit. We recognise the incongruence of this period with previous works and since we also do not sample a full rotation period, the period and amplitude should be considered to be minimum estimates.

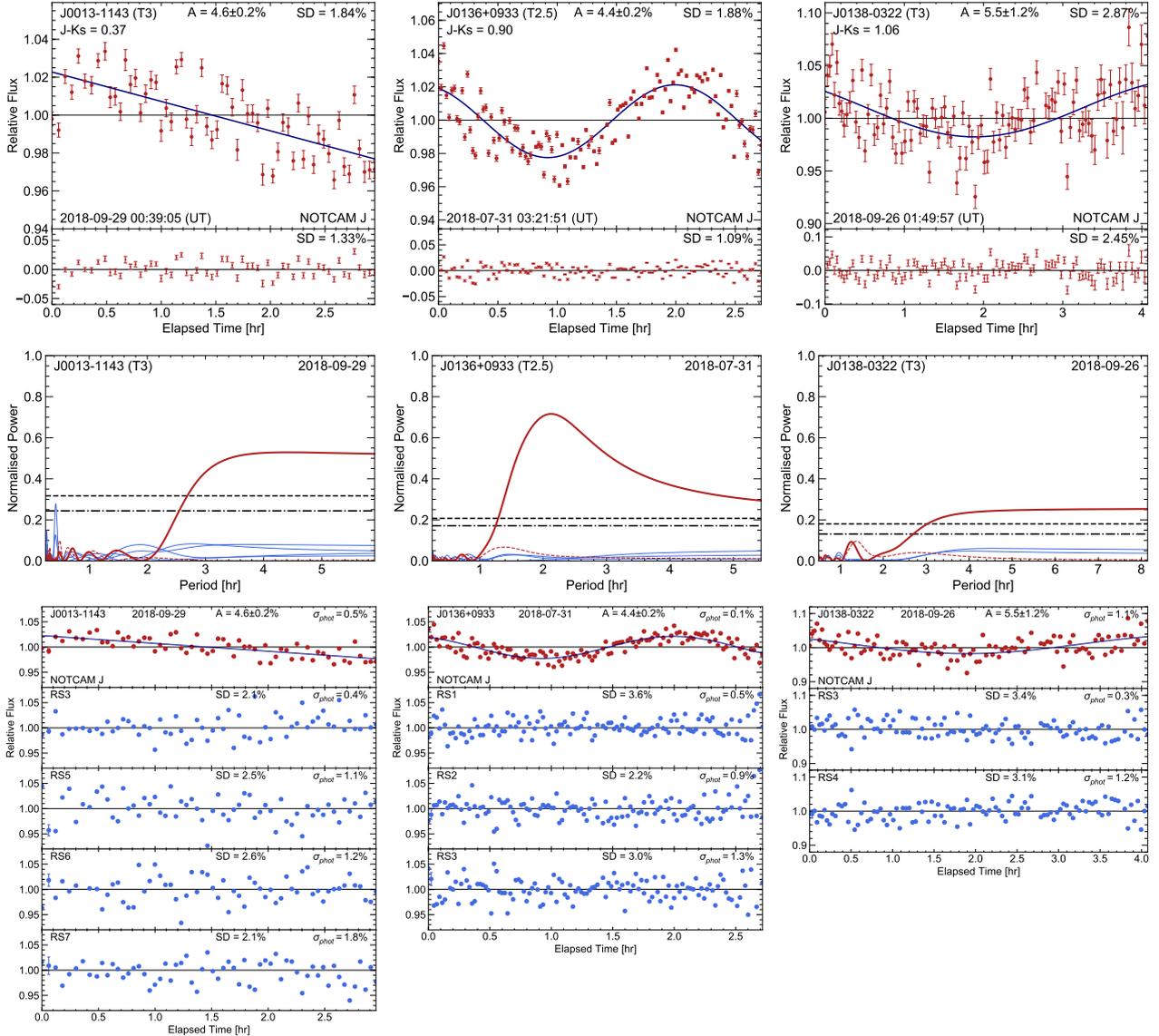

**Fig. 3.** Plots for detections of significant variability in the observations of J0013 in 2018 (*left set* of three figures), J0136 in 2018 (*middle set*) and J0138 in 2018 (*right set*). (I) For a given set of figures, the *top figure* consisting of two panels shows the final light curve with photometric uncertainties (red circles with error bars) with the best fit (dark blue solid line). The peak-to-peak amplitude and uncertainty from the fit along with the standard deviation (SD) of the light curve are also listed. *Bottom panel*: residual light curve (data-fit) and its SD. (II) The *middle figure* shows the generalised Lomb-Scargle periodogram for the target (thick solid red line), reference stars (thin blue lines) and the target residual (dashed red line). The horizontal upper dashed, lower dotted and dashed lines represent the 0.1% and 1.0% false alarm probability levels respectively. (III) In the *bottom figure*, the top panel shows the target light curve (red circles), the best fit (dark blue solid line). The remaining panels show the light curves of the reference stars (RS) used to create the median combined reference trend. The mean photometric uncertainties for the target and RS are also indicated by the error bar on the second data point. The Bayesian periodograms for the first two observations are available in Fig. 2.

### 4.1.3. 2MASS J01383648−0322181

Originally classified as the T3 BD WISEPC J013836.59−032221.2 by Kirkpatrick et al. (2011), J0138−0322 was classified as T2.5 by Robert et al. (2016), but as this was done using low-quality spectra we keep the original T3 classification from Kirkpatrick et al. (2011). First observed in visitor mode during somewhat poor conditions on the night of 2017-10-11, J0138−0322 has been one of the more problematic targets to analyse. The field of view suffers from a severe lack of good and stable reference stars, with two bright stars that saturate at integration times optimal for this 16.36 magnitude (*J*) BD, which simultaneously represents an insufficient amount of integration time for the two very faint stars. Due to bad weather, this first epoch resulted in only 1.5 h of usable observations which in turn are sub-optimal in quality. Using all four available RS, we detect very strong but non-significant variability for this brief epoch with a LM fit amplitude of $8.9 \pm 1.6\%$ and obtain a FAP of 5% from the GLSP, seen together with the light curves in Appendix B. This could qualify it as a marginal detection but we note that all RS share weaker, but similar power in the periodogram even if they do not show the same pattern of variability in the light curves, leading to a non-significant classification for this epoch. It did however make the target a priority for follow-up in 2018.

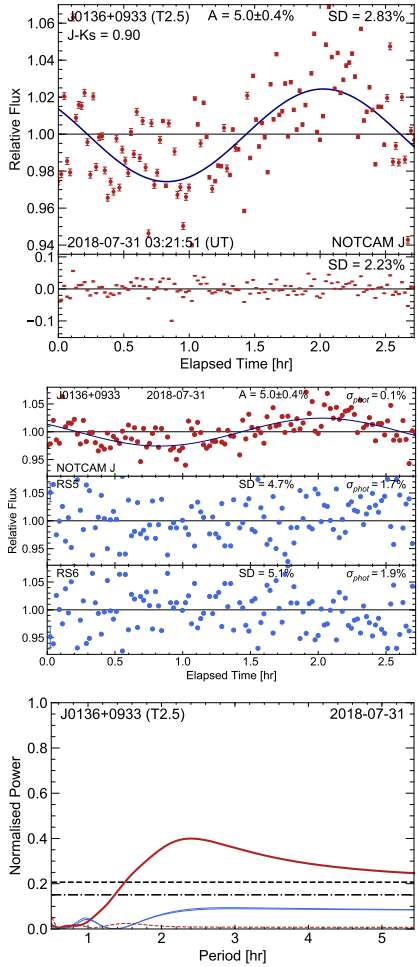

**Fig. 4.** Same type of plots as for J0136 in Fig. 3, here illustrating the robustness of differential photometry by showing the result from only using two of the worst available RS. With only two, their light curves will be near perfect reflections of each other, but the target light curve is still remarkably well corrected. This supports the conclusion that the variability detected in the observation of J0138−0322 (2018) is significant, despite a lack of available RS.

For the next epoch an attempt was made to shift the field slightly to a more favourable configuration of references, with only marginal success. The observing conditions on 2018-09-25 were much improved from 2017 for most of the night until morning twilight, and J0138 was observed continuously for four hours with six RS present in the field. Unfortunately, two of these saturated during periods of excellent seeing and two were much too faint with photometric uncertainties of ∼3−4%. Fortunately, the two remaining ones were stable with one being of similar brightness as the target, with the resulting GLSP and light curves seen in Fig. 3 (right column). For this epoch we detect highly significant strong variability with a minimum amplitude of $5.5 \pm 1.2\%$ and a period of at least three hours. As we appear to be centred in the trough of the light curve, the measured amplitude and rotation are difficult to evaluate in relation to a possible true amplitude and rotation period. If the true period is close to the observation baseline, the light curve seen in the previous epoch could be explained without the need for a double-peaked light curve.

The lack of more than three good references forced us to rely on only two RS, which merits a brief discussion on the reliability of such an analysis. If only two RS are used, then the light curve of one reference will mirror the appearance of the other. However, this only marginally affects the reliability of an analysis of the target light curve. To illustrate this we present in Fig. 4 the GLSP and light curves from using only the two worst references available to the J0136 observation. The obtained LM fit is very similar to the one presented previously for J0136, even though the SD of the residual has more than doubled. The peak-period indicated by the GLSP has increased somewhat, and while the power is lower overall, the curve rests comfortably well above the FAP level. Similar tests can be done with other targets, indicating the same level of reliability. In short, using only two references can lead to a slight under- or overestimation of the amplitude due to increased scatter while periodogram power tends to be underestimated. If the references are also of good quality, the difference between two and three RS is even smaller. Both references for this observation are well behaved, so we would expect the result to change only marginally and remain a significant detection if we had more of them available.

### 4.1.4. 2MASS J22153705+2110554

First discovered by Kellogg et al. (2015) in the SDSS data and classified as a weak T1 (T0 + T2) binary candidate, satisfying two of the criteria from Burgasser et al. (2010a), J2215+2110 was selected for this survey from Robert et al. (2016). We observed J2215 in visitor mode over two epochs, first under cloudy conditions during the night of 2017-10-13 for 2.5 h, followed by 2018-09-25 with excellent conditions for five hours. Strong and highly significant variability was detected for both epochs (left and middle columns in Fig. 5).

For the final analysis of the first epoch, four RS of similar brightness were selected from a total of 16 viable RS in the field. Through LM fitting we measure a peak-to-peak amplitude of $10.7 \pm 0.4\%$, which remains stable within ∼10−12% even with poorer references. As this observation was somewhat more affected by changes in flux due to dithering, the chosen RS also exhibit this effect to some extent for the purpose of correcting for it in the target, which as can be seen in the GLSP was highly successful. The non-significant peak of the residual is likely induced by the RS and not due to intrinsic rotational modulation. From the fit we measure a period of $3.0 \pm 0.2$ h, which is in agreement with the GLSP which indicates a peak period at around three hours. As we do not fully sample the probable peak at the end of the observation, period and amplitude measurements should be considered minimum values for this epoch.

For the second epoch, six of 15 RS were initially selected for the final analysis but reducing this to the three most similar in brightness slightly improved the fit statistics. From the LM fit we measure a peak-to-peak amplitude of $2.6 \pm 0.2\%$ and a rotational period of $5.2 \pm 0.5$ h, which is in agreement with the peak periods of $5.7 \pm 0.7$ and 5.7 h obtained from the BGLSP (bottom left in Fig. 2) and GLSP (Fig. 5) respectively. We had not been expecting such an interesting and drastic reduction of 3/4 in amplitude but, as will be discussed in depth in Sect. 5, this is emerging as a relatively common phenomena. On-site analysis of the observation in $J$ did however prompt a shift in observing strategy as we had planned a night of $K_S$ observations, but since one can typically only expect an amplitude ratio of $A_{K_S}/A_J \sim 0.25$–$0.50$ (e.g. Radigan et al. 2012; Vos et al. 2019) other targets took priority.

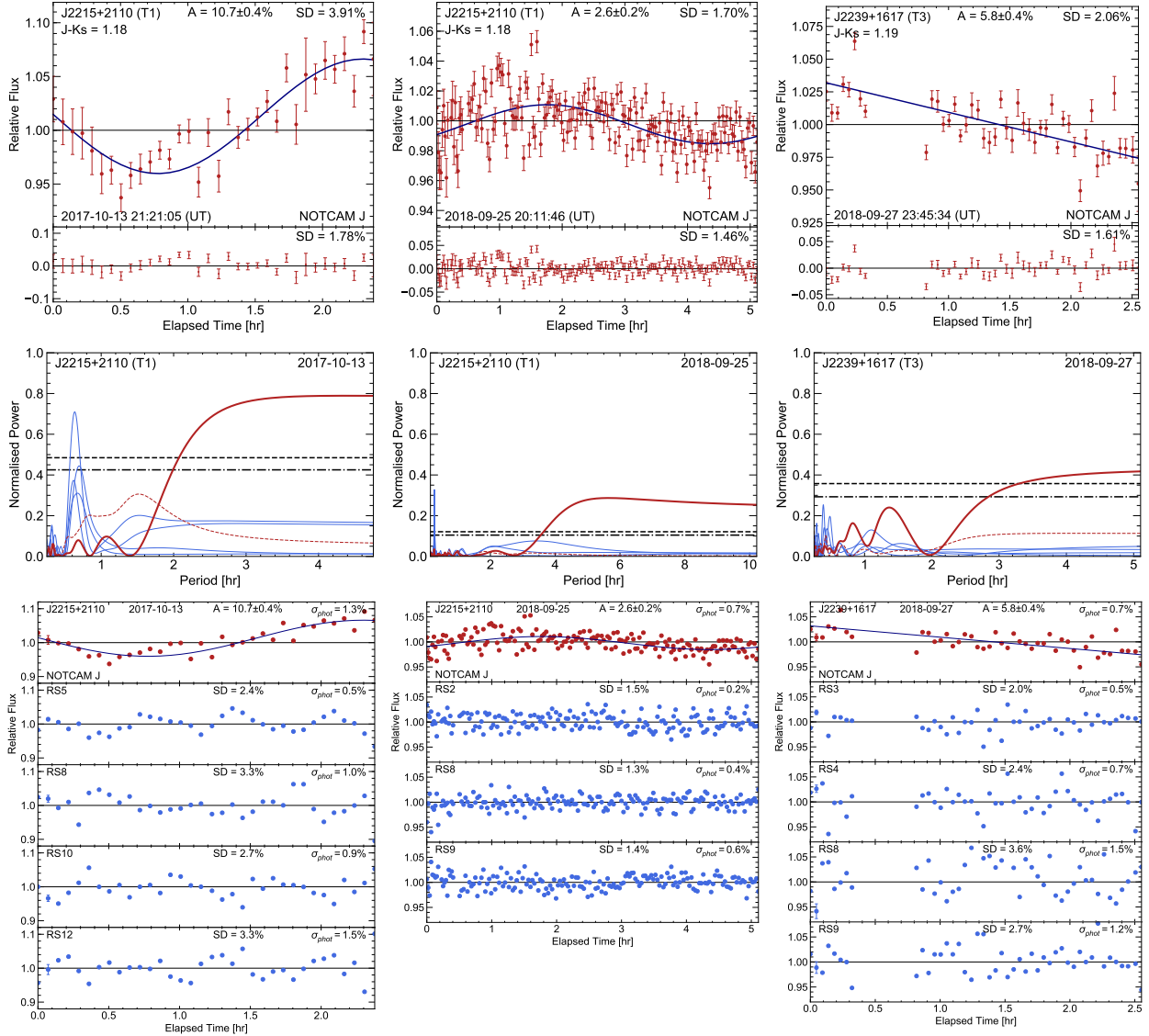

**Fig. 5.** Same as Fig. 3 but for the detections of significant variability in the observations of J2215 in 2017 (*left set* of three figures), in 2018 (*middle set*) and J2239 in 2018 (*right set*). The Bayesian periodogram for J2215 in 2018 is available in Fig. 2.

Our findings for J2215+2110 represent an important addition to the T-dwarfs currently known to be very strongly variable, second only in $J$ peak-to-peak amplitude to J2139+0220 for a given epoch (Table D.1; Radigan et al. 2012; Apai et al. 2013). This level of variability, combined with the drastic long-term amplitude evolution, makes J2215 especially interesting for future multi-wavelength observations. These could investigate amplitude ratios (Sect. 5.1.2) and wavelength dependent phase shifts (e.g. Yang et al. 2016), two important aspects that help probe the 3D structure of the atmosphere and the underlying physical mechanisms. As half of our new strong and significant variables (J0013 and J2215) were also previously flagged for potential unresolved binarity, we discuss this aspect further in Sect. 5.1.3.

### 4.1.5. 2MASS J22393718+1617127

This T3 BD, selected from Robert et al. (2016), was like J0138−0322 originally discovered by Kirkpatrick et al. (2011), as WISEPC J223937.55+161716.2. We observed J2239+1617 during cloudy conditions on the night of 2018-09-27 for 2.5 h. Despite the somewhat poor conditions, effectively corrected for during the differential photometry process as previously discussed in Sect. 3.1 and illustrated in Fig. 1, we confidently detect significant variability (FAP < 0.1%) in the shape of a linear trend with amplitude $5.8 \pm 0.4\%$, as seen in the right side panels in Fig. 5. The half-hour gap early in the light curve was caused by thicker clouds but could be giving rise to a discrepancy between the GLSP and BGLSP. The non-significant peak (FAP $\sim 10\%$) at around 1.5 h in the GLSP is not indicated in the BGLSP,

**Table 3.** Detections of significant and non-significant variability for all epochs in this work.

| Target | Epoch (UT) | Amplitude [a] (%) | Period [b] (h) | FAP (%) | $\Delta t$ (h) | $\sigma_{BD}$ [c] (%) | $\sigma_{RS}$ [c] (%) | Trend |
|---|---|---|---|---|---|---|---|---|
| Detections (FAP ≤ 0.1%) | | | | | | | | |
| J0013−1143 | 2018-09-28 | 4.6 ± 0.2 | 2.8 | ≪0.1 | 2.95 | 0.5 | 1.1 | Linear |
| J0136+0933 | 2018-07-30 | 4.4 ± 0.2 | 2.13 ± 0.02 | ≪0.1 | 2.76 | 0.1 | 0.9 | Sinusoidal |
| J0138−0322 | 2018-09-25 | 5.5 ± 1.2 | 3.2 | ≪0.1 | 4.07 | 1.1 | 0.8 | Sinusoidal |
| J2215+2110 | 2017-10-13 | 10.7 ± 0.4 | 3.0 ± 0.2 | ≪0.1 | 2.40 | 1.3 | 1.0 | Sinusoidal |
| J2215+2110 | 2018-09-25 | 2.6 ± 0.2 | 5.2 ± 0.5 | ≪0.1 | 5.15 | 0.7 | 0.4 | Sinusoidal |
| J2239+1617 | 2018-09-27 | 5.8 ± 0.4 | 3.4 | <0.1 | 2.64 | 0.7 | 1.0 | Linear |
| Marginal detection | | | | | | | | |
| J2148+2239 [d] | 2018-09-26 | 2.1 ± 0.3 | 2.4 ± 0.4 | 1.9 | 4.19 | 0.8 | 0.7 | Sinusoidal |
| Non-significant variability | | | | | | | | |
| J0135+0205 | 2018-09-28 | <1.4 ± 0.4 | … | … | 2.44 | 0.7 | 0.5 | Linear |
| J0138−0322 | 2017-10-11 | <8.9 ± 1.6 | … | ~5 | 1.55 | 0.6 | 0.7 | Sinusoidal |
| J0150+3827 | 2018-09-26 | <1.9 ± 0.3 | … | ~25 | 1.72 | 0.5 | 0.6 | Linear |
| J0316+2650 | 2018-09-26 | <1.1 ± 0.3 | … | … | 2.42 | 0.6 | 0.7 | Linear |
| J2132−1452 | 2017-10-11 | <1.5 ± 0.3 | … | … | 4.12 | 1.5 | 1.0 | Linear |
| J2132−1452 | 2018-09-27 | <1.7 ± 1.1 | … | … | 1.11 | 1.1 | 0.5 | Linear |
| J2148+2239 | 2017-10-12 | <4.4 ± 1.6 | … | ~30 | 0.84 | 1.5 | 0.5 | Linear |
| J2303+3150 | 2018-09-28 | <3.6 ± 0.3 | … | ~70 | 1.64 | 0.6 | 0.3 | Sinusoidal |

**Notes.** Resulting variability detections from all epochs and epochs. Detections for targets other than J0136 represent first detections. Detailed information on NIR spectral type, colour and photometric distances, with references, can be found in Table 1. Our significant detections are also included in the Colour ($J − K_s$) vs SpT diagram in Fig. 7. [a]"Peak-to-peak" or "peak-to-trough" amplitudes obtained from Levenberg-Marquardt best fits. In general, for most observations with unconstrained periods, these should be considered minimum amplitudes as the full rotation period was not covered. [b]Given the sometimes rapid light curve evolution of BDs, see e.g. J1324+6358 in Yang et al. (2016), Apai et al. (2017) or J0136+0933 in Artigau et al. (2009), Croll et al. (2016), tabulated periods should be considered to be minimum periods unless well constrained from multiple epochs. Period estimates with uncertainties were obtained from the sinusoidal fit, with the remainder representing where the GLSP crossed the 0.1% false alarm probability (FAP) level. [c]Median photometric uncertainty for the target ($\sigma_{BD}$) and the average of the median photometric uncertainties for the reference stars used in the final result ($\sigma_{RS}$). [d]With a FAP of 1.9% and ten targets, we would expect ~0.2 false positives in the survey and as such regard J2148 to be a highly likely candidate for strong variability. However, for a more robust comparison with the surveys of Radigan et al. (2014) and Vos et al. (2019) we exclude it from the final statistics.

and could be due to the gap. Both peridograms favour a period greater than four hours, and the non-significant peaks in the GLSP are also eliminated from the residual. The curve crosses the 0.1% FAP at 3.4 h so this defines our minimum period for this epoch. With a minimum amplitude of 5.4% over just 2.5 h, J2239 could, along with J0013, possibly reach peak-to-peak amplitudes at levels similar to those measured in 2017 for J2215.

### 4.2. Marginal detection of variability in 2MASS J21483578+2239427

Discovered by Robert et al. (2016) as the T1 BD SIMP J21483578+2239427, J2148+2239 was observed during both epochs in visitor mode and is our only target with marginally detected variability at a FAP of 1.9%. The first epoch of 2017-10-12 was severely affected by clouds, leaving only 45 min of high-noise data from a total of three hours. The light curve for J2148 in Appendix B shows a linear trend of 4.4 ± 1.6% and the GLSP indicates a FAP of ~30% for the longer period non dither-related peak. Not much more could be concluded from this observation due to the extremely short baseline and poor data quality, and J2148 was followed-up during the next epoch.

We returned to J2148 on the night of 2018-09-26 under much better conditions for 4.2 h. The field of view for J2148 is very rich in references and the final five RS used in Fig. 6 were selected from a pool of 24 RS. From a LM fit to the light curve we measure a peak-to-peak amplitude of 2.1 ± 0.3% and a period of 2.4 ± 0.4 h. The GLSP and BGLSP (bottom right in Fig. 2) indicate peak periods of 2.5 and 2.5 ± 0.3 h respectively. The clear periodicity is easily discerned by eye and present in all RS iterations, but still falls short of the 0.1% FAP in the GLSP at 1.9%. A possible reason for this is the rather significant peak at the dithering period, which in this case was not successfully corrected for by differential photometry as the RS similarly affected by dithering were also very variable (at longer periods). The added noise from this signal combined with a seemingly very rapid rotation could explain why we fail to reach the level of significance one would expect from a by eye estimate. For a rotation period of ~2.4 h, the longest linear trend one would observe is around one hour. While the 2017 epoch observation is not of high enough quality or length to help confirm the period measured for this epoch, it does not exclude it.

These findings indicate that J2148 could have a nearly identical rotation period to that of J0136, while being strongly variable and of similar SpT. Future multi-wavelength observations of J2148 could therefore allow for a direct comparison between the two objects, shedding light on differences or similarities in the short-term evolution of variability.

### 4.3. Non-significant detections of variability

Using a similar structure as in Sect. 4.1, we briefly summarise the analysis of our non-significant detections. All plots detailing

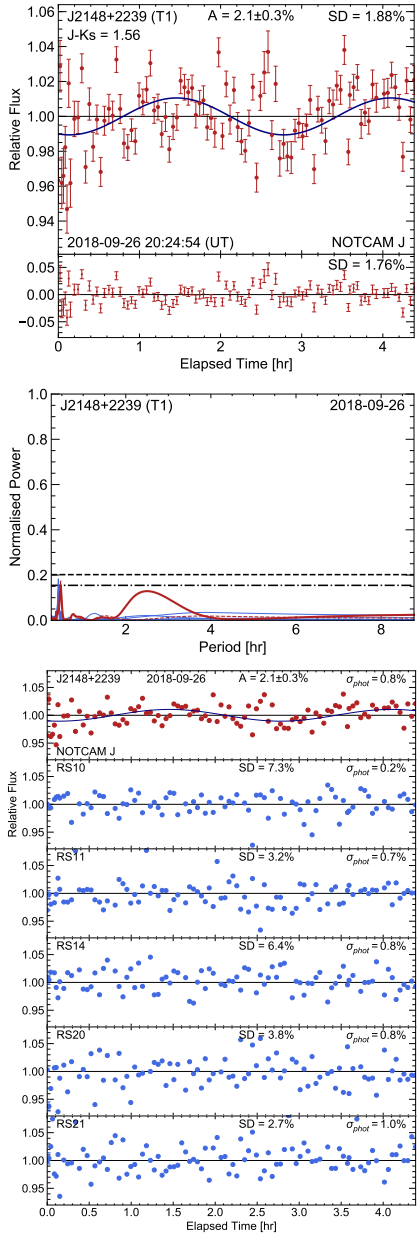

**Fig. 6.** Same as Figs. 3 and 5, but for the detection of marginally significant variability (FAP of 1.9%) during the observation of J2148 in 2018. The Bayesian periodogram is available in Fig. 2.

light curves and GLSP for these non-significant detections are available in Appendix B.

### 4.3.1. 2MASS J01352531+0205232

This L9.5 BD was first discovered by Best et al. (2015) as PSO J023.8557+02.0884 and was added to the survey from Robert et al. (2016). J0135+0205 is our faintest target at 16.62 in $J$ and was observed in visitor mode for 2.5 h at the end of the 2018-09-28 night under mainly good conditions. Rapidly increasing airmass and approaching twilight led to a decreased photometric precision for the last hour of the observation. From the resulting GLSP and light curves we obtain a non-significant linear trend amplitude of $1.4 \pm 0.4\%$ that is shared by several RS in the field.

### 4.3.2. 2MASS J01500997+3827259

First discovered by Robert et al. (2016) as the L9.5 BD SIMP J01500997+3827259, J0150+3827 was observed during the night of 2018-09-26 in visitor mode for 1.7 h under good conditions. The light curve obtained with five out of 17 RS shows a linear trend with a best-fit amplitude of $1.9 \pm 0.3\%$ and a corresponding non-significant peak-power FAP of ~25% in the GLSP. This variability seems limited to the target and not common to the RS (unlike J0135) so we therefore consider it tentatively variable.

### 4.3.3. 2MASS J03162759+2650277

Discovered by Best et al. (2015) as the T2.5 BD PSO J 049.1159+26.8409, and rated as a strong T2 + T7.5 binary candidate (4/6 criteria), J0316+2650 was included from Robert et al. (2016). After attempted observations in 2017, J0316 was observed for 2.5 h on 2018-09-27 under excellent conditions. We note that J0316 was still affected by dithering effects after differential photometry using five out of 16 RS, but do not expect this to conceal any significant variability in the linear trend, as the GLSP power is close to zero.

### 4.3.4. 2MASS J21324898−1452544

First discovered as a T3 BD by Andrei et al. (2011), J2132−1452 was included as SpT T3.5 from Robert et al. (2016). J2132 is a very low visibility target at the NOT, and we obtained a good baseline of 4.1 h in the early evening on 2017-10-11 during partly cloudy conditions and high airmass. We attempted a follow-up on 2018-09-27 which was cut short by rain, leaving 1.1 h of average quality data. Neither epoch shows any signs of significant variability, with a GLSP power near zero, using any set of eight possible RS.

### 4.3.5. 2MASS J23032925+3150210

Discovered by Schneider et al. (2016a) as the T2 BD WISEA J230329.45+315022.7, we included J2303+3150 as a T3 from Robert et al. (2016) and observed it for 1.6 h before clouds became too severe. During this short baseline J2303 displays strong sinusoidal variability but at a FAP level of 70%. Both the GLSP and LM fit favour a period around 1.4–1.5 h, a very rapid rotation shared only by the strongly variable and well studied BD J2228−4310 (Table D.1), but this could simply be a spurious signal due to poor observing conditions.

## 5. Discussion

In this section we first compare our findings with those of a number of other surveys that included BDs in our focussed spectral range of L9–T3.5 (Girardin et al. 2013; Metchev et al. 2015; Radigan et al. 2014) and present our results in this wider context in a Colour vs SpT diagram (Fig. 7). Next we present a catalogue (Tables 4 and D.1) of currently known strong variables to be found in the literature, containing information on all relevant observation epochs we were able to locate. This is followed by a discussion of the broader implications of our results along with a number of interesting aspects of unresolved binarity, long-term evolution of variability and amplitude ratios. Finally we take a closer look at our findings for J2215+2110.

Out of the ten targets surveyed for variability for the first time in this work, four out of the ten are strongly variable ($A > 2\%$) at a high significance level (>99%). In addition to this, J0136+0205 is still strongly variable, but excluded from this discussion of first-discoveries. These are all presented in the Colour vs SpT diagram in Fig. 7 as red circles alongside other strong variables identified in the literature (discussed in Sect. 5.1 and catalogued in Tables 4 and D.1), showing the sizeable contribution of this work to the number of known strong variables in the L/T transition. Using small number Poisson statistics from Gehrels (1986) for our four detections out of the ten possible, we infer a probability distribution for the fraction of strongly variable L9–T3.5 BDs of $40^{+32}_{-19}$%.

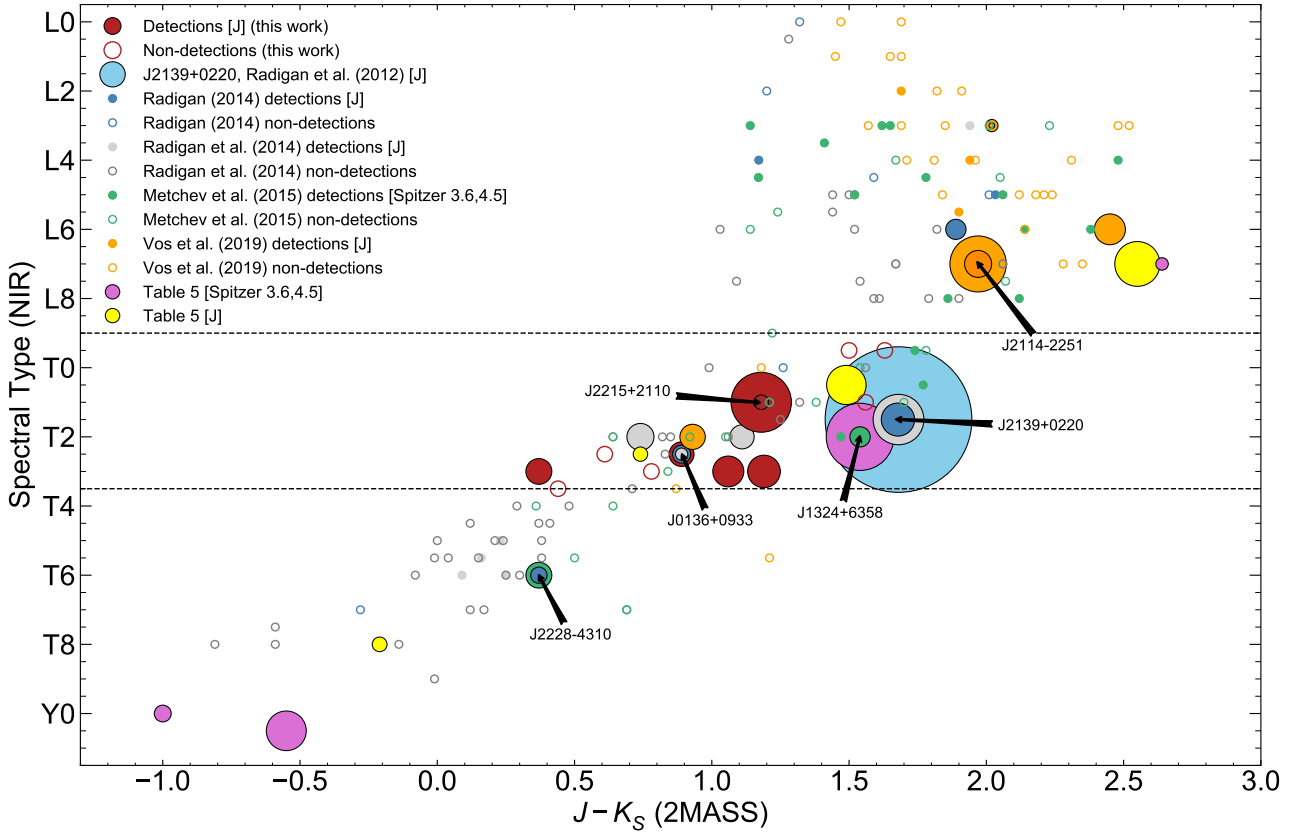

**Fig. 7.** Colour $(J−K_S)$ vs Spectral Type diagram showing detections (filled circles) and non-detections (empty circles) of variability from numerous sources in the literature, including the strong variables listed in Table D.1. Circle size is proportional to the peak-to-peak amplitude, with the smallest filled circles indicating <2% and larger ones indicating strong variability, ranging from 2 to 26%. Detections from this work, representing a significant addition to known strongly variable T-dwarfs, and non-detections are shown in dark red filled and larger empty circles respectively, all 11 present within the L9–T3.5 L/T transition indicated by the two dashed lines. The three large variability surveys done by Radigan et al. (2014) (grey circles), Metchev et al. (2015) (green) and Vos et al. (2019) (orange) included 62, 44 and 30 BDs respectively. The independent analysis of the Wilson et al. (2014) survey by Radigan et al. (2014) (teal circles) included another 14. The largest amplitude observed by Radigan et al. (2012) is indicated by the largest (light blue) circle. Pink and yellow filled circles represent a number of other strong variables from Table D.1.

The survey of Girardin et al. (2013) was similar in scope with observations of nine targets from the ground in $J$-band, finding one out of five L9–T3.5 BDs in their sample to be strongly variable. Peak-to-peak amplitudes of 2–3% were measured for the T0.5 SDSS J105213.51+442255.7, which after their observations was resolved as a binary by Liu et al. (2010), making it one of few strongly variable BDs that are also resolved binaries, with the most well studied one being the nearby (2 pc) WISE J104915.57−531906.1AB ("Luhman 16AB", e.g. Luhman 2013; Gillon et al. 2013).

Metchev et al. (2015) surveyed 44 L3–T8 BDs, of which 11 non-binary targets were L9–T3.5 L/T transition objects, using the *Spitzer* telescope at 3.6 and 4.5 μm. They find two of these to be strong variables, with lower-amplitude variability detected in a few others. The most notable detection was the T2 2MASS J13243559+6358284 ($A = 3.05$%), which we will return to several times in this section as it was later extensively studied with *Spitzer* by Yang et al. (2016), Apai et al. (2017), yielding results highly relevant to our detections. While Metchev et al. (2015) do not consider a statistical analysis limited only to the L/T transition, they find that $36^{+26}_{-17}$% of T0–T8 BDs are variable with $A > 0.4$% and that nearly all BDs are variable to some degree at the 0.2% level.

The most extensive and detailed survey of L/T transition objects to date was made by Radigan et al. (2014) during their large-scale ground based survey of 62 L4–T9 targets in the $J$-band. Their survey results were also the premise for the survey presented in this work, as strong variability was absent in all 41 targets outside the L/T transition, while 4/16 inside were significantly variable at peak-to-peak amplitudes of 3–9%, with the most well known being J0136+0933 and 2MASS J21392676+0220226. Their statistical analysis resulted in a probability distribution which indicated that $39^{+16}_{-14}$% of L9–T3.5 BDs are variable at the >2% level, a figure very much consistent with our own inferred distribution of $40^{+32}_{-19}$%.

Radigan et al. (2014) further investigate the effects of inclination and explore possible correlations between $J − K_S$ colour and variability, finding that $80^{+18}_{-19}$% of L9–T3.5 BDs with $0.8 > J − K_S > 1.5$ should be variable when viewed edge-on.

These possible correlations were previously suggested by both Kirkpatrick et al. (2010) and Metchev et al. (2015), but investigated in depth by Vos et al. (2017) for the majority of the known variable BDs. By determining the $v \sin i$ and calculating the inclinations of 19 targets, they found clear correlations with variability amplitude for both inclination and $J - K_S$ colour anomaly (i.e. how much the colour deviates from the median for a given SpT). They draw two main conclusions relevant to this work. Firstly, $J$-band variability at any level is only detected in targets inclined at $>35°$, with the strongest known variable J2139+0220 ($A = 26\%$ Radigan et al. 2012) being inclined edge-on at $i = 90°{}^{+0}_{-1}$ and other strong variables also showing high inclination values, indicating a correlation between amplitude and target inclination. Secondly, the correlation between $J - K_S$ colour anomaly and amplitude comes out naturally as there is a correlation between inclination and colour anomaly, where bluer BDs are viewed closer to pole-on and consequently redder are viewed closer to equator-on. Since we find a substantial number of new strong variables with peak-to-peak amplitudes of $>4.5\%$, and one of these also stands out as blue (J0013−1143), we search for similar connections between variability and colour anomaly in our own sample.

To obtain a median from which to estimate the $J - K_S$ colour anomaly ($\Delta_{J-K_S}$) for our targets, a criteria query[5] was run on the Simbad database to collect all non-binary BDs for a given SpT. The resulting Fig. 8 includes 203 L8–T5 BDs from the database, all our targets and J2139+0220 for reference as it was shown by Vos et al. (2017) to be maximally inclined. Even considering uncertainties in SpT of up to ±1.5, J2139 ($\Delta_{J-K_S} = -0.61$) would be significantly redder than a typical BD of the same type. The T3s J0138 ($\Delta_{J-K_S} = -0.50$) and J2239 ($\Delta_{J-K_S} = -0.63$) show similarly strong anomalies, along with the T1 J2148 ($\Delta_{J-K_S} = -0.44$), potentially supporting the idea that it is strongly variable. J2215 ($\Delta_{J-K_S} = -0.06$) on the other hand appears to be of average colour, which could indicate that we are not viewing it near edge-on. J0013 ($\Delta_{J-K_S} = 0.19$) is indeed unusually blue, but with substantially larger errors than our other targets it is hard to draw any conclusions. Photometry for J0136 ($\Delta_{J-K_S} = -0.15$) is well constrained, putting it comfortably towards the red, and Vos et al. (2017) report an inclination of $i = 80 \pm 12°$. None of our non-detections appear to show significant anomalies towards the blue or the red, and while the T3.5 J2132 ($\Delta_{J-K_S} = -0.32$) in this plot is unusually red, there are few BDs in that SpT bin to establish a robust median. Judging from the overall trend of the median colours of SpT T2–T4, J2132 is likely average. To conclude, other than J0013, all our targets with significant or marginal variability have colours in the interval of $0.8 > J - K_S > 1.5$ suggested by Radigan et al. (2014) as the most probable for variability (if viewed edge-on). As we now add a significant number of strong variables to the literature, obtaining $v \sin i$ measurements of these new variables should help strengthen correlations between colour, inclination and variability.

## 5.1. Cataloguing strong variability in the literature

As the collective knowledge in this field has grown substantially in the past few years, the need for a comprehensive variability catalogue arises. Vos et al. (2017) include a catalogue of the majority of all known variable BDs with period estimates, including spectral type, single-epoch variability amplitude in $J$

---

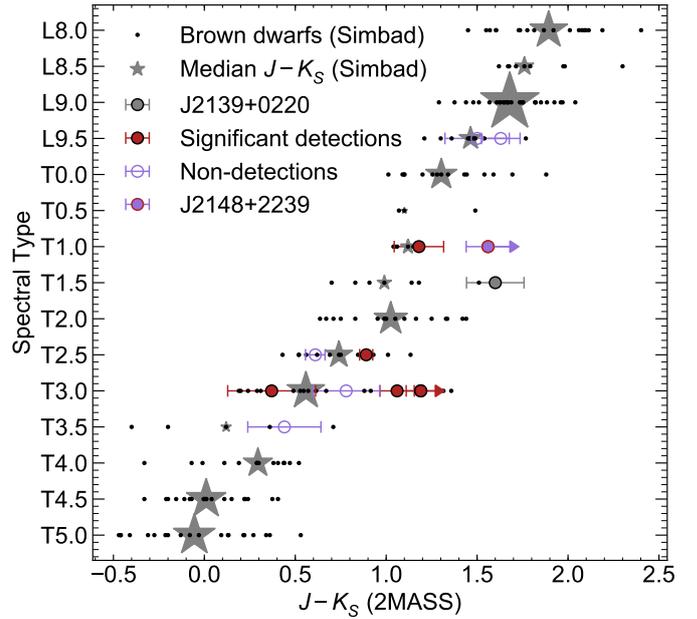

**Fig. 8.** Colour ($J - K_S$) vs Spectral Type diagram illustrating distribution of our targets in terms of colour anomaly. Black points show 203 brown dwarfs in the L/T transition gathered by criteria from the SIMBAD database for various SpT, with the median colour shown as a grey star for a given SpT (size proportional to the number of BDs in that bin). J2139+0220 is included for reference as it is very strongly variable and has been shown to inclined at 90°. Detections and non-detections are shown with $J - K_S$ error bars in dark red filled and purple empty circles respectively, with the marginal detection of J2148 shown as a combination of the two. Poorly constrained 2MASS $K_S$ photometry for J2148 and J2239 is indicated by upper limit arrows.

and *Spitzer* 3.6 and 4.5 μm. With recent dedicated multi-epoch studies (e.g. Yang et al. 2016; Apai et al. 2017) probing the evolution of variability on both short and long timescales, continued follow-ups of known variables and the seemingly ever clearer trend of finding strong variability in the L/T transition with a high probability, we find that the creation of a catalogue dedicated to strong variables including as much multi-epoch information as possible is well motivated. Strongly variable BDs are the most well suited for studying the underlying atmospheric physics and chemistry giving rise to variability in these objects, as this requires simultaneous multi-wavelength observations to probe the atmosphere in 3D. As peak-to-peak amplitudes can change drastically from one band to another (e.g. $A_K/A_J \sim 0.5$, Sect. 2.2) this will tend to exclude weak variables from effective multi-wavelength study. As an example of this, the T6 J2228−4310 is variable at half the amplitude in $J$ than at 3.6 μm ($A_{3.6}/A_J \sim 2.0$), while the opposite relation applies for T2.5 J0136+0933 where $A_J/A_{3.6} \sim 3.3$ (Yang et al. 2016).

The catalogue consists of an object information table providing discovery, SpT, magnitudes and distance information with references in Table 4 followed by Table D.1 cataloguing all known epochs we were able to track down for an object having shown strong variability $A > 2\%$ in $J$, $H$, $K$, 3.6 and 4.5 μm at any given epoch. In this table we catalogue SpT, $J - K_S$, epoch, peak-to-peak amplitude, period, observational baseline, telescope and instrumental information, the latter illustrating a broad range of ground-based telescopes that have been used to study variability in BDs, highlighting the usefulness of similar telescopes for future surveys. In the case where a study spanned a large number of consecutive nights, for example the 17 ground based nights

---

[5] Criteria for Simbad query, e.g.: "sptype = T1.5 & otype = BD* & $J$ mag <20 & $K$ mag <20", giving all non-binary BDs for SpT T1.5 with $J$ and $K$ magnitudes in the database. Survey targets were excluded.

**Table 4.** Detailed object information for strong variables currently in the literature.

| 2MASS ID | Alternate ID | SpT (Opt.) | Ref. | SpT[a] (NIR) | Ref. | 2MASS $J$ (mag) | 2MASS $K_S$ (mag) | $J - K_S$ (mag) | $d_\pi$ (pc) | Ref. |
|---|---|---|---|---|---|---|---|---|---|---|
| 2MASS J00470038+6803543 | WISEP J004701.06+680352.1 | L7 | 19 | L5 | 23 | 15.17 | 12.62 | 2.55 | 12.2 ± 0.3 | 23 |
| 2MASS J01365662+0933473 | SIMP J013656.57+093347.3 | ... | ... | T2.5 | 2 | 13.45 | 12.56 | 0.89 | 6.14 ± 0.04 | 18 |
| 2MASS J00473039−1216155 | PSO J071.8769−12.2713 | ... | ... | T2 | 4 | 16.48 | 15.55 | 0.93 | 27.7 ± 2.9[c] | 4 |
| 2MASS J05012406−0010452 | WISEA J050124.21−001047.1 | L3 | 11 | L4 | 3 | 14.98 | 12.96 | 2.02 | 20.7 ± 0.6 | 23 |
| 2MASS J07584037+3247245 | SDSS J075840.33+324723.4 | T2 | 3 | T2 (T0+T3.5?) | 6 | 14.95 | 13.88 | 1.07 | 16 ± 2 | 16 |
| ... | WISE J085510.83−071442.5 | ... | ... | >Y2 | 21 | 25[b] | ... | ... | 2.20 ± 0.24 | 26 |
| 2MASS J10101480−0406499 | WISEA J101014.55−040649.9 | ... | ... | L6 | 10 | 15.51 | 13.62 | 1.89 | 16.7 ± 2.6 | 17 |
| 2MASS J10491891−5319100 B[d] | WISE J104915.57−531906.1 B | ... | ... | T0.5 | 8 | 11.22[b] | 9.73[b] | 1.49[b] | 2.02 ± 0.15 | 25 |
| 2MASS J10521350+4422559 | SDSS J105213.51+442255.7 | ... | ... | T0.5 (L6.5+T1.5) | 15 | 15.96 | 14.57 | 1.39 | 26.1 ± 0.5 | 15 |
| 2MASS J11472421−2040204 | WISEA J114724.10−204021.3 | ... | ... | L7 | 29 | >17.51 | 14.87 | >2.64 | 31.2 ± 1.5[c] | 29 |
| 2MASS J12073346−3932539 b | 2M1207b | ... | ... | L5−L8 | 9 | >18.5 | 16.93 | >1.57 | 54.2 ± 1.1 | 14 |
| 2MASS J13004466+1222325 C | Ross 458C | ... | ... | T8 | 7 | 16.67 | 16.88 | −0.21 | 11.7 ± 0.2 | 30 |
| 2MASS J13243553+6358281 | WISEA J132434.93+635827.4 | ... | ... | T2 (L8.0+T3.5?) | 6 | 15.60 | 14.06 | 1.54 | 9 ± 1 | 32 |
| ... | WISEP J140518.40+553421.5 | ... | ... | Y0.5 | 12 | 21.06[b] | 21.61[b] | −0.55[b] | 6.93 ± 0.24 | 27 |
| 2MASS J15164806+3053443 | SDSS J151643.01+305344.4 | ... | ... | T0.5 (L8+L9.5?) | 6 | 16.77 | 15.16 | 1.61 | 23 ± 3 | 16 |
| 2MASS J16291840+0335371 | SIMP J162918.41+035537.0 | ... | ... | T2 | 13 | 15.29 | 14.04 | 1.25 | 15 ± 3[c] | 13 |
| ... | WISEP J173835.52+273258.9 | ... | ... | Y0 | 12 | 19.58[b] | 20.58[b] | −1.00[b] | 7.34 ± 0.44 | 27 |
| 2MASS J21140802−2251358 | PSO J318.5338−22.8603 | ... | ... | L7 | 22 | 16.71 | 14.74 | 1.97 | 22.2 ± 0.9 | 23 |
| 2MASS J21392676+0220226 | WISEA J213927.09+022023.9 | T0 | 28 | T1.5 (L8.5+T3.5?) | 28 | 14.71 | 13.58 | 1.13 | 9.9 ± 0.2 | 31 |
| 2MASS J21443131+1446190 B | HN Peg B | ... | ... | T2.5 | 24 | 15.86[b] | 15.12[b] | 0.74 | 17.9 ± 0.2 | 23 |
| 2MASS J22282889−4310262 | WISEA J222829.01−431029.8 | ... | ... | T6 | 5 | 15.66 | 15.3 | 0.36 | 10.6 ± 0.9 | 17 |
| 2MASS J22443167+2043433 | WISEA J224431.89+204340.2 | L6.5 | 20 | L6 | 1 | 16.48 | 14.02 | 2.45 | 17.0 ± 0.3 | 23 |

**Notes.** Catalogue of known strong variables currently available in the literature, identified at high levels of significance, in descending 2MASS ID with spectral, colour and parallactic distance information. [a]Possible binarity indicated in parenthesis, with J1052 being the only resolved binary (Dupuy et al. 2015). [b]Magnitudes given in MKO instead of 2MASS. [c]Estimated photometric distance. [d]The secondary component of the resolved, nearby binary system Luhman 16AB (L7.5 primary component), with MKO magnitudes obtained from the SpT reference (8).

**References.** (1) Allers & Liu (2013); (2) Artigau et al. (2006); (3) Bardalez Gagliuffi et al. (2015); (4) Best et al. (2015); (5) Burgasser et al. (2006); (6) Burgasser et al. (2010a); (7) Burgasser et al. (2010b); (8) Burgasser et al. (2013); (9) Chauvin et al. (2004); (10) Cruz et al. (2003); (11) Cruz et al. (2009); (12) Cushing et al. (2011); (13) Deacon et al. (2011); (14) Ducourant et al. (2008); (15) Dupuy et al. (2015); (16) Faherty et al. (2009); (17) Faherty et al. (2012); (18) Gagné et al. (2017); (19) Gizis et al. (2015); (20) Kirkpatrick et al. (2008); (21) Leggett et al. (2015); (22) Liu et al. (2013); (23) Liu et al. (2016); (24) Luhman et al. (2007); (25) Luhman (2013); (26) Luhman (2014); (27) Martin et al. (2018); (28) Reid et al. (2008); (29) Schneider et al. (2016b); (30) Scholz (2010); (31) Smart et al. (2013); (32) Theissen (2018).

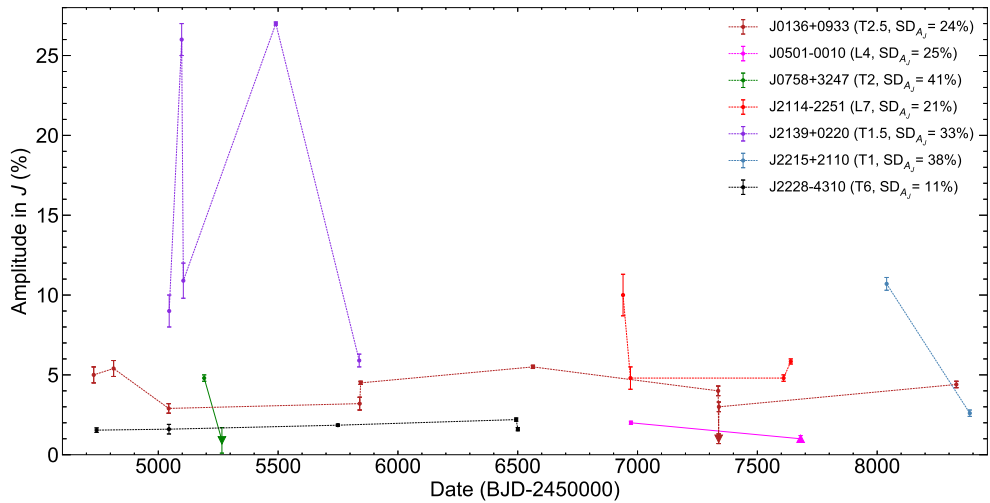

**Fig. 9.** Peak-to-peak amplitudes in $J$ against BJD−2450000, illustrating the long-term evolution over several epochs for a number of strong variables in the literature with multi-epoch observations. The first epoch plotted is for J0136 in 2008-09-21, amplitude uncertainties are indicated by error bars and epochs with upper/lower limits are plotted with triangles. One object, the T6 J2228 stands out as being relatively stable, indicated in the legend by $SD_{A_J}$, the standard deviation (SD) for a given object obtained from the SD of all epochs normalised by the maximum amplitude.

for J0136+0933 in Croll et al. (2016) or the large *Spitzer* telescope programme of Yang et al. (2016), Apai et al. (2017), we report the most reliable and relevant epochs. It could be argued that objects exhibiting variability at amplitudes of ∼1% in 3.6, 4.5 µm should be included as potentially strong variables, given that a number of BDs show (much) larger amplitudes in $J$ than in the mid-infrared (MIR). However, as exemplified earlier by the $A_J/A_{3.6}$ ratios for J0136 and J2228, this is not always true.

Our aim for this catalogue is to aid further research in the field by providing a comprehensive overview of the current observational database on strong variables. This effectively highlights a number of emerging and interesting trends that we choose to focus more closely on in the sections below. One is the long-term evolution of variability that is just now being more extensively studied and while it is too early to claim any significant correlations, we discuss some potential trends in Sect. 5.1.1. Another is the value of multi-wavelength observations (Sect. 5.1.2) and the importance of simultaneity for establishing reliable amplitude ratios. Finally, the possible correlation between objects flagged for unresolved binary and strong variability, discussed in a growing number of works, could be useful in selecting targets for future surveys (Sect. 5.1.3).

### 5.1.1. Long-term evolution of variability

Focus in the field has so far, for natural reasons, been on discovering new variables and studying short-term variability. This was most recently done in depth by Yang et al. (2016), Apai et al. (2017) where they obtained over a thousand hours of space based observations with *Spitzer* and the HST during 2012–2014. Now, with an increasingly large number of discoveries and individual epochs being added to the literature, it is also starting to become possible to probe the long-term evolution of these objects. Figure 9 was created using data from the catalogue in Table D.1 and depicts the long-term evolution of the peak-to-peak amplitude in $J$ for seven targets, with four of these having four or more epochs. J2139+0220 (T1.5) stands out visually as the most "chaotically" variable with rapid shifts in amplitude over relatively short periods of time, while J2228−4310 (T6) in context seems unusually quiescent, showing minor relative changes over five years. J0136 has been observed for just under ten years and is remarkably consistent with a median amplitude of 4.2%.

It would thus appear that some BDs such as J2139 and J2215 fluctuate more than others (J0136, J2114 and J2228) in relative peak-to-peak amplitude. The minimum amplitude measured over five epochs in $J$ for J2139 is 5.9% while the maximum is 27%, giving a ratio between the two of 0.22. For our two epochs of J2215, the amplitude changes drastically giving a correspondingly similar ratio of 0.24. The strongly variable T2 J1324+6358 is not included in the figure, as it has only been observed in the MIR, but has shown extreme short-term amplitude evolution in the 2–12% range, potentially corresponding to $J$-band amplitudes as high as J2139 with similar ratios. For J2114, J0136 (excluding the upper limit estimate of 1% on 2015-11-12) and J2228 the ratios are 0.48, 0.53 and 0.73 respectively. J2228 was until very recently the only known strong variable among late T-dwarfs. Previously surveyed by Metchev et al. (2015), the planetary-mass T8-dwarf Ross 458C has now been shown to be variable at 2.62% in $J$ by Manjavacas et al. (2019b), with early indications of possible similarities with J2228. As more BDs with strong variability are identified, especially among later T-dwarfs, and monitored over several epochs, a search for correlations between spectral type and long-term amplitude evolution should be feasible. Probing such variability trends as a function of spectral type is a promising avenue for future studies, and may enhance our understanding of the underlying mechanisms. Another related correlation to examine would be the volatility and frequency with which the amplitude changes, indicated in Fig. 9 by a standard deviation measure.

### 5.1.2. Amplitude ratios from multi-wavelength observations

It was showed early on (e.g. Artigau et al. 2009; Buenzli et al. 2012; Radigan et al. 2012; Apai et al. 2013; Biller et al. 2013) that multi-wavelength observations are critical in constraining the physical mechanisms giving rise to observed variability in BDs, as different layers of the atmosphere are probed at different wavelengths. Several surveys and studies over the years include amplitude ratios, most frequently $A_K/A_J$ due to the prevalence of ground-based observations, but few of these are the result of simultaneous observations. The most common practice is near-simultaneous, achieved by observing a few hours in $J$ followed by a few in $K$ which for ground-based observations typically results in short baselines in both bands. If variability in BDs arise from single spots this is not necessarily an issue, but the large and apparently rapid variability fluctuations discussed in the previous section imply a more complex mechanism, suggesting that observations need to be simultaneous to obtain reliable ratios. For ground-based observations in the NIR, there are few options for simultaneous observations other than swapping back and forth between filters, applied successfully by for example Vos et al. (2019) for J2114−2251 during the 2016 observations using the NTT, and Radigan et al. (2012) for J2139+0220. The latter also showed a significant drift towards lower $A_K/A_J$, but not $A_H/A_J$, over the course of a week, indicating that the ratios likely also are subject to short-term evolution. *Spitzer* and the HST have been combined by a number of programmes with good results (e.g Buenzli et al. 2012; Yang et al. 2016; Apai et al. 2017; Biller et al. 2018) to give simultaneous $3.6\mu\mathrm{m}/A_J$, $A_{4.5}/A_J$ or near-simultaneous $A_{4.5}/A_{3.6}$ ratios for a number of objects, with J2228−4310 having the largest amount of combined epochs available. Using these we give an example of how this type of information could potentially be used to better understand the long-term evolution of variability, as the number of multi-wavelength epochs increase.

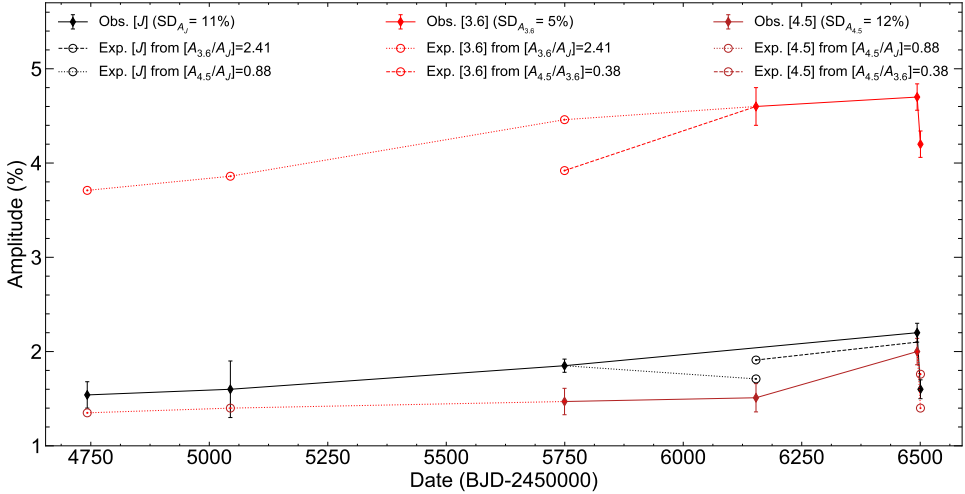

**Fig. 10.** Peak-to-peak amplitudes against BJD−2450000, showing the amplitude evolution for the T6 J2228−4310 in $J$ (black points), 3.6 $\mu$m (bright red) and 4.5 $\mu$m (dark red). Observed amplitudes are indicated by solid lines and calculated expected amplitudes (based on empiric amplitude ratios) by empty circles with dashed or dotted lines. By using empiric amplitude ratios obtained from simultaneous observations (given in legend), one can calculate an expected amplitude at a given epoch for a given filter. This process is described in detail in Sect. 5.1.2.

In Fig. 10 we present all available epochs for J2228 over five years from Table D.1 and utilise amplitude ratios obtained from these to estimate the peak-to-peak amplitudes in all three bands at all epochs. To illustrate the process, one can from the last two epochs in 3.6 $\mu$m and $J$ separated by seven days obtain an average ratio $A_{3.6}/A_J = 2.41 \pm 0.22$ with an uncertainty estimated by the standard deviation. $A_{4.5}/A_J = 0.88 \pm 0.07$ is similarly obtained from the two epochs separated by two years. Combining these we can calculate the expected amplitudes in $J$ at Barycentric Julian Date (BJD) 2456153, yielding $A_J = 1.91\%$ from the observed $A_{3.6} = 4.7\%$ or $A_J = 1.71\%$ from $A_{4.5} = 1.51\%$. Similarly we can calculate expected amplitudes for 3.6 $\mu$m at BJD 2455749 using amplitude ratios based on $J$ and 4.5 $\mu$m observations. This can in principle be extended to comparing observed to computed amplitudes at other epochs, although this leads to more permutations than are practical to illustrate in Fig. 10. The exercise nonetheless implies that amplitudes in adjacent wavelength bands can be predicted with reasonable (but not perfect) accuracy, highlighting the importance of continued multi-wavelength monitoring.

If amplitude ratios themselves are subject to short-term evolution over time as shown for $A_K/A_J$ in Radigan et al. (2012), these types of predictions would be more unreliable. It is not yet clear however if this is the case for all amplitude ratios, and in the event only certain ratios are prone to such evolution, that in and of itself could be a useful tool. For example, J0136 was shown by Yang et al. (2016) to not exhibit the same phase shift between HST $J$ and *Spitzer* Ch1 and Ch2 as has been observed in other targets such as J2228 by Buenzli et al. (2012),

Yang et al. (2016). In this case, both authors also measure the phase shift for J2228 between HST $J$ and *Spitzer* Ch2 to be 118 ± 7° and 156.5 ± 9.5° respectively, with the difference possibly explained by a separate patchy cloud deck at lower pressure levels Yang et al. (2016). From the light curves, amplitude ratios of $A_{4.5}/A_J \approx 0.80$ and $A_{4.5}/A_J \approx 0.95$ can also be obtained, potentially indicating an increasing ratio with increasing phase shift in this one case. So while amplitude ratios might be unreliable for some bands, they could still prove useful in the search for possible correlations.

Finally, it is important to note that for long-period variables, near-simultaneous *Spitzer* observations likely give more reliable amplitude ratios than short-period ones, for example J2228 in Metchev et al. (2015). Over their 21 h observation, ten full rotations pass in 3.6 $\mu$m before five are observed in 4.5 $\mu$m and as has been shown for instance Buenzli et al. (2015); Croll et al. (2016); Apai et al. (2017), significant changes can occur between or even during rotations indicating that caution should be used when basing an amplitude ratio estimate on the median or average of a several full rotations. A better alternative might be to use the rotation closest in time to the shift in bands, assuming there is a clear periodic signal without significant evolution.

### 5.1.3. Unresolved binarity or strong variability

For variability studies like the one presented in this work, potential contamination by unresolved binaries presents two main problems. Firstly, any detected variability can not be attributed to a specific component, as it can arise in either one or both. Secondly, any observed amplitude will be diluted if the variable component is significantly fainter than the primary in the observed band. Radigan et al. (2014) roughly estimate that 15−30% of L9–T3.5 BDs are unresolved binaries at separations of ≳2−3 AU, based on previous binary fraction studies, translating to ∼1−3 potential unresolved binaries in our target sample. In the cases of J0013 (T3) and J2215 (T1) which were both flagged for potential unresolved binarity (Kellogg et al. 2017, 2015), the suggested individual components of T3.5 + T4.5 and T0 + T2 are similar enough that any dilution of the observed amplitude is likely to be marginal. For J0316 the suggested components are T2 + T7.5 (Best et al. 2015), in which case a variable amplitude from the T7.5 component could be suppressed by the brighter T2. The baseline for J0316 is too short (∼2.5 h) for any definitive conclusions to be drawn concerning a correlation between unresolved binarity and variability. Strong variability of ∼2% over a long rotation period, as has been found for HN Peg B ($P \sim 15.4$ h; Zhou et al. 2018), would likely yield a linear amplitude of $\sim 0.3-0.5\%$ over 2.5 h and thus not be detectable above the noise during our observation baseline. Further epochs with longer baselines are needed to determine if J0316 follows the same trend as J0013 or J2215.

Resolved binaries are generally excluded from variability surveys for the aforementioned problems, and as Radigan (2014) notes, L/T transition binaries often only have one transition object, exacerbating the problem. To date the only resolved binary where both components have been studied for variability is the J1049−5319 AB system (L7.5 + T0.5 with 3.1 AU separation; Burgasser et al. 2013), with the B component included in the catalogue. While the resolved and unresolved observations by Biller et al. (2013) were separated by a week, they found resolved amplitudes for J1049 B to be two to three times greater, with a non-detection for J1049 A, when compared with earlier unresolved observations of J1049 AB. Unresolved binaries on the other hand should not automatically be excluded, even if the suggested individual components have radically different SpT. As part of their search for composite atmospheres in their HST WFC3 spectral library of 76 BDs, Manjavacas et al. (2019a) find that the spectral indices used by Burgasser et al. (2006, 2010a) and Bardalez Gagliuffi et al. (2014) to identify unresolved binaries are potentially biased towards variable BDs, suggesting they might be useful for selecting candidates for variability studies. It is interesting to note that J0136 was also flagged as a strong candidate for binarity during their search according to these indices. As far as we can judge from the literature, J0136 has not previously been considered an unresolved binary, possibly because it was noted by Artigau et al. (2009) that binarity had been excluded down to $0.2''$ (200 mas, $\sim 1.2$ AU). It nonetheless fits in well with the growing trend among strong variables in our catalogue, where several of the strongest variables, other than our two new discoveries, have also been flagged for unresolved binarity by the same index criteria.

The earliest indication of this connection was seen with J2139+0220, initially thought to be an unresolved binary but later shown by Radigan et al. (2012) to be very strongly variable, with binarity ruled out down to a maximum separation of <130 mas (1.6 AU) for a contrast difference in $J$ of <3 mag. The potential trend was then highlighted by Metchev et al. (2015) with the discovery that J1324+6358 was strongly variable and flagged for unresolved binarity. In the literature we also find J0758+3247 and J1516+3053 reported as strong variables, and candidates for unresolved binarity based on these criteria. Bardalez Gagliuffi et al. (2015) observe 43 M- and T-dwarfs, 17 with suggested binarity from spectral indices, and exclude binarity for J0758+3247 and J1516+3053 down to <65 mas (<1.0 AU) and <70 mas (<2.7 AU) respectively. Initially thought to be a single BD J1052+4422 was surveyed for variability but later confirmed to be a resolved binary (70 mas, 1.84 AU; Dupuy et al. 2015), making it the only one in the catalogue other than J1049 B. No high-resolution imaging results exist yet for J1324, but Gagné et al. (2018) determine that it is likely a member of the AB Doradus young moving group and show signs of low surface gravity, arguing that this feature sufficiently explains the observed composite atmosphere spectra that leads to it being classified as an unresolved binary. Low surface gravity has previously also been associated with an increased fraction of (strong) variability in BDs (e.g. Biller et al. 2018; Vos et al. 2019), and interestingly, Gagné et al. (2017) also recently identified J0136 as a member of the Carina-Near young moving group. For J1324 Gagné et al. (2018) are able to weakly constrain the presence of an equal-mass or equal-luminosity binary down to 17.3 mas ($\sim 0.22$ AU) and 200 mas (2.5 AU) respectively, and finally conclude that rather than being over-luminous, as expected for binaries, it is under-luminous by $\sim 0.4$ mag compared to field dwarfs.

In summary, it seems unlikely that any of the strong variables presently flagged for unresolved binarity in our catalogue are actual binaries, which would strengthen the trend. Further observations with for instance high-resolution imaging should be done of J0013 and J2215 to help verify the possibility that these spectral index criteria mistakenly characterise strongly variable BDs as unresolved binaries.

### 5.2. J2215+2110: Conclusions

J2215+2110 is an especially important addition to the group of variable T-dwarfs that have been extensively studied in other works and discussed in the previous sections. Its T1 (Kellogg et al. 2015) or T1.5 (Robert et al. 2016) SpT classifications places it comfortably in the middle of the L/T transition and the 2MASS photometry has relatively small uncertainties, resulting in a reliable $J - K_S$ colour with Fig. 8 indicating it is slightly redder than the median colour of either SpT T1 or T1.5. The light curves presented in Fig. 5 show different periods at each epoch, with the almost twice as long period (5.2 vs 3.0 h) from the second epoch indicating a double-peaked light curve in the first. This is not an unusual feature, as we explored for J0136 in Sect. 4, and appears from the studies of Yang et al. (2016), Apai et al. (2017) to be common among the strongest variables. The full light curves of their extensive *Spitzer* observations are available in the supplementary material to Apai et al. (2017), showing varying double peaked features for J0136, J1324 and J2139. Both J0136 ($P = 2.4$ h) and J1324 ($P = 13$ h) show a double peaked light curve with a period of almost exactly $P/2$, persisting throughout the whole observation for four full rotations (9.6 and 52 h). While we lack light curves covering J2215 for a full rotation in either epoch, the estimated periods suggest that $P_1/P_2 \sim 1/2$, indicating a similar light curve behaviour as seen for J0136 and J1324. With this similarity in mind, the true rotation period for J2215 is likely ≤6 h. This similarity also suggests that the underlying mechanism for variability in J2215 is more complex than what would be expected from a single spot model, thus requiring an approach similar to Apai et al. (2017) where both spots and patchy cloud bands with varying brightness are modelled. This combined effect on the variability from patchy clouds and spots can also be clearly seen in great detail for Jupiter in the work by Ge et al. (2019).

One final conclusion that we draw for J2215 from the catalogue concerns the minimum and maximum amplitudes one might expect to observe in the future, given the data available in the literature on strong variability. If the relation between $A_{min}/A_{max}$ is $\sim 1/4-1/6$ (Sect. 5.1.1), we would therefore from the first epoch not expect to observe a significantly lower amplitude in $J$ than $\sim 1.8-2.7\%$ at any other epoch and conversely not a larger amplitude than $\sim 10-15\%$ based on the second epoch. The continued study of J2215 is well positioned to make clear contributions to our understanding of L/T transition objects, joined by both J0013 and J2239 as potentially very strong variables.

## 6. Summary

We have surveyed ten brown dwarfs in the L9–T3.5 spectral type range for first discoveries of periodic variability. For four of these we report significant detections with peak-to-peak amplitudes of 2.6−10.7%, representing a substantial addition to currently known strong variables in the L/T transition. Two of these targets had previously been flagged for unresolved binarity due to signs of composite atmospheres in their spectra, and the detection of strong variability could indicate that their composite appearance instead could be due to rotationally modulated variability. These discoveries add to the growing number of strong variables that have potentially been similarly misclassified, indicating that spectra criteria used for identifying unresolved binarity could aid in target selection for future surveys.

We also present another epoch of the previously known strong variable J0136+2650 (Artigau et al. 2009), confirming that it is still variable three years after the last reported observation with a similar amplitude, giving further indication of the long-lived nature of variability in brown dwarfs. With the addition of this epoch, J0136 has now been regularly observed for almost a decade, making it unique among variable brown dwarfs. Furthermore, marginally significant (FAP 1.9%) strong variability was detected for J2148+2239, indicating a very similar rotation period to that of J0136. If confirmed from future epochs, this would allow for a direct comparison between two brown dwarfs of similar spectral type, variability amplitude and near-identical rotation periods. Such a comparative study could highlight possible correlations between rotation rate and variability or, since J0136 is likely part of a young moving group (Gagné et al. 2017), differences in how variability evolves or is modulated in young objects. Excluding this marginal detection, we infer a variability occurrence fraction among L9–T3.5 dwarfs of $40^{+32}_{-19}\%$, in good agreement with the occurrence fraction of $39^{+16}_{-14}\%$ reported by Radigan et al. (2014).

*Acknowledgements.* Based on observations made with the Nordic Optical Telescope, operated by the Nordic Optical Telescope Scientific Association at the Observatorio del Roque de los Muchachos, La Palma, Spain, of the Instituto de Astrofísica de Canarias. We thank the anonymous referee for a very well written and useful report that improved this manuscript. We thank all the staff at the ORM and NOT for their support, especially senior staff astronomer Anlaug Amanda Djupvik for her assistance during and after the 2018 observations. We further acknowledge the helpful comments from Evan O'Connor, Markus Janson gratefully acknowledges support from the Knut and Alice Wallenberg Foundation. This survey made use of the CDS services SIMBAD and VizieR, as well as the SAO/NASA ADS service and the archival services for the HST and SST.

# Appendix A: Global observing condition plots for all epochs

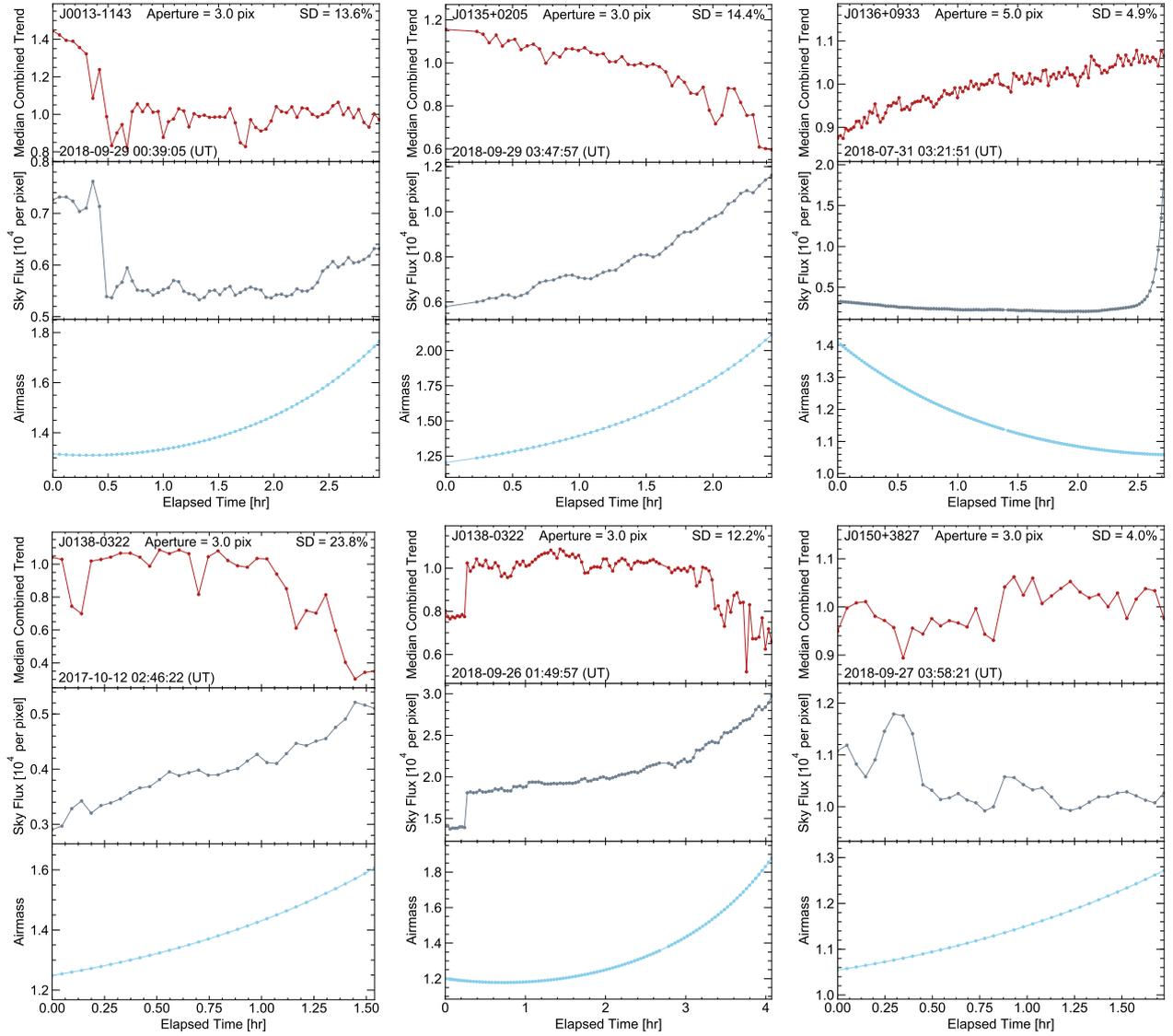

**Fig. A.1.** Three-panel figures for six epochs illustrating the quality of the local observing conditions. (I) *Top panel*, in a given figure: (normalised) median combined reference trend obtained from the selected set of reference stars for a given observation. Also listed is the chosen aperture size (in pixels) and the standard deviation of the trend curve. (II) *Middle panel*: background sky flux per pixel. (III) *Bottom panel*: airmass evolution and gives an easy overview of the number and frequency of discarded or lost frames.

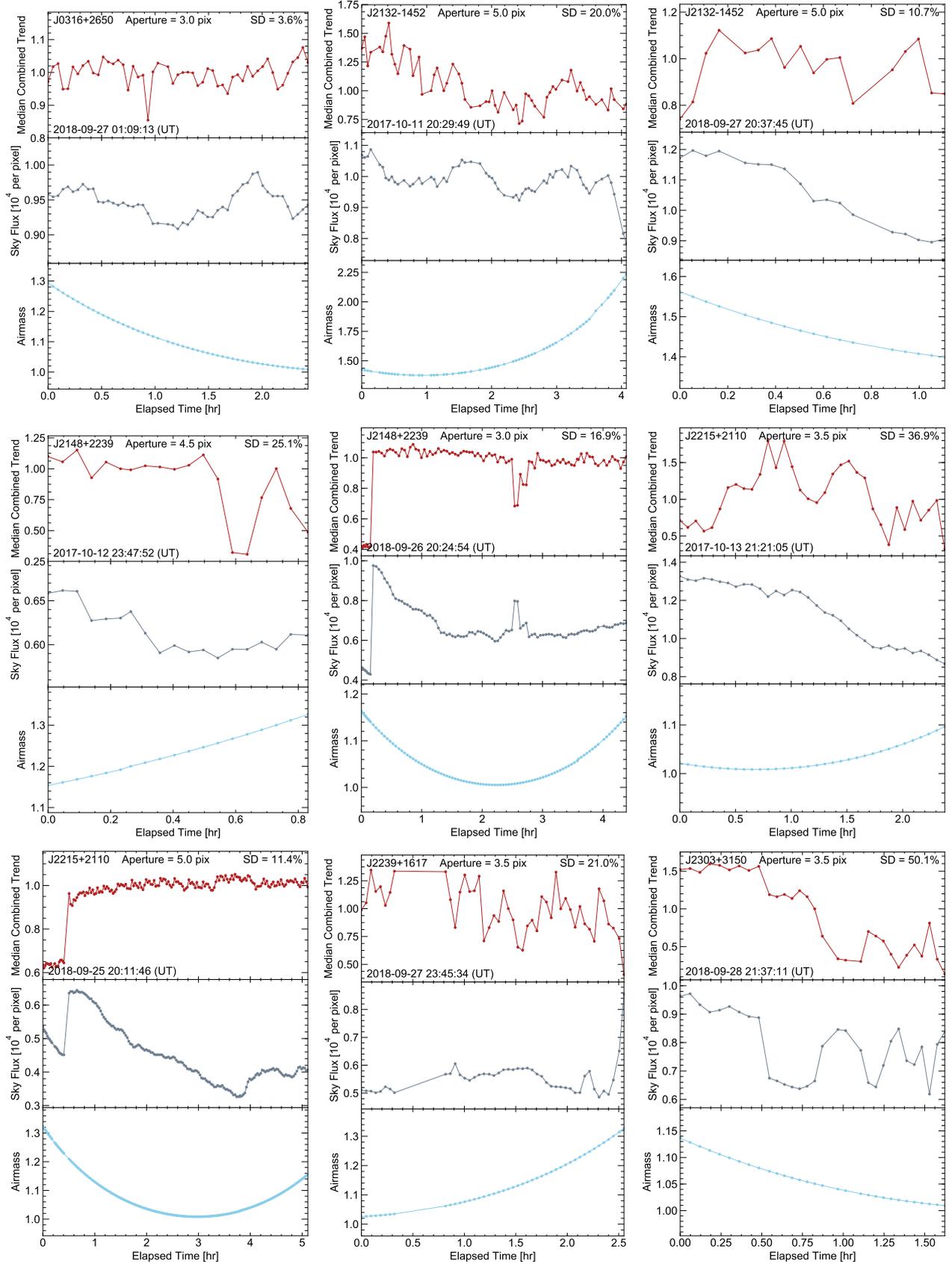

**Fig. A.2.** Same as Fig. A.1 for the remaining nine epochs. The sudden increase in flux seen early on in e.g. 2148 (2018) and J2215 (2018) is due to an increased integration time.

## Appendix B: Light curves for non-significant detections

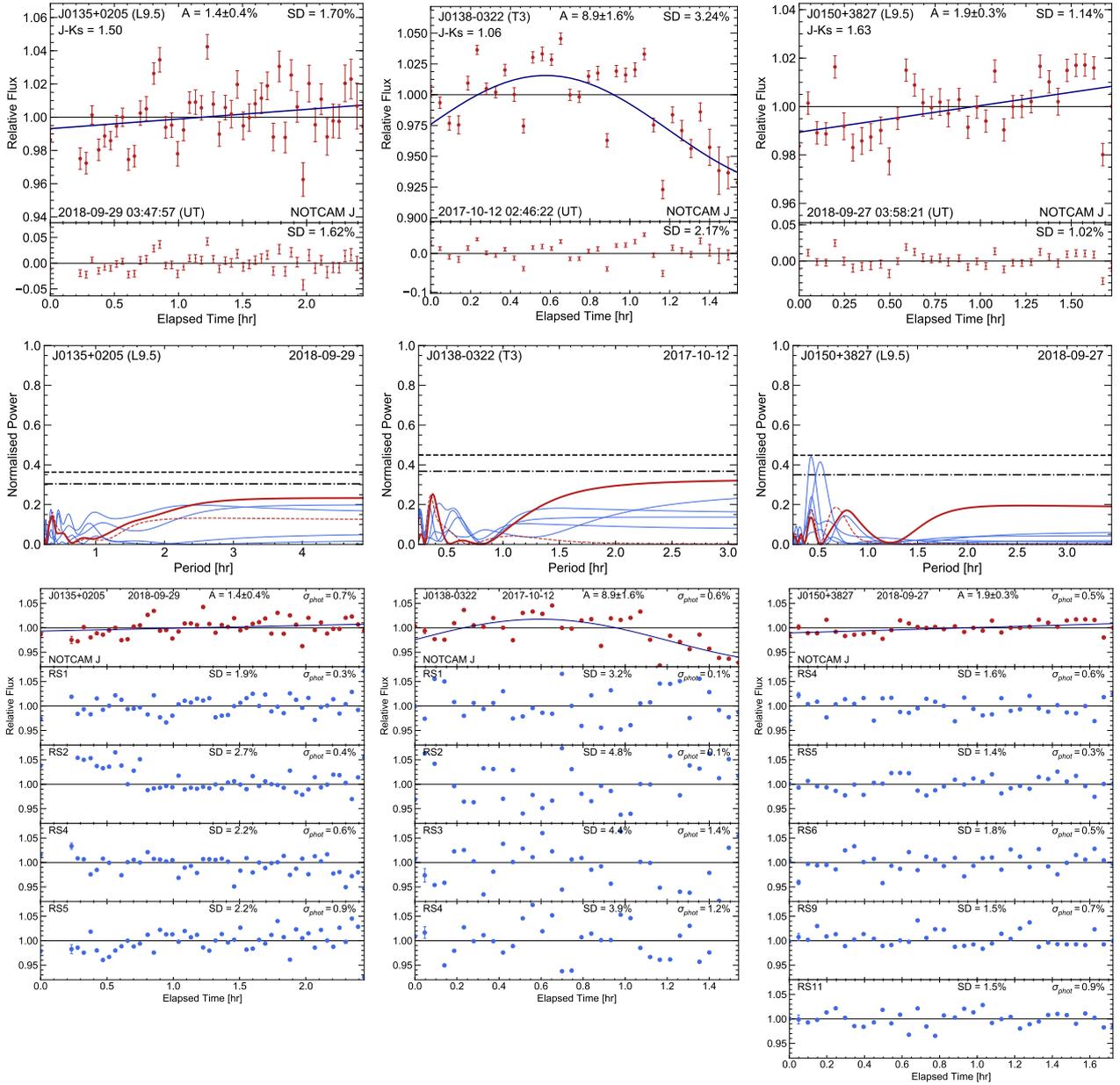

**Fig. B.1.** Plots for the non-detection observations of J0135 in 2017 (*left set* of three figures), J0138 in 2017 (*middle set*) and J0150 in 2018 (*right set*). (I) For a given set of figures, the *top figure* consisting of two panels shows the final light curve with photometric uncertainties (red circles with error bars) with the best fit (dark blue solid line). The peak-to-peak amplitude and uncertainty from the fit along with the standard deviation (SD) of the light curve are also listed. The bottom panel shows the residual light curve (data-fit) its SD. (II) The *middle figure* shows the generalised Lomb-Scargle periodogram for the target (thick solid red line), reference stars (thin blue lines) and the target residual (dashed red line). The horizontal upper dashed, lower dotted and dashed lines represent the 0.1% and 1.0% false alarm probability levels respectively. (III) In the *bottom figure*, the top panel shows the target light curve (red circles), the best fit (dark blue solid line). The remaining panels show the light curves of the reference stars (RS) used to create the median combined reference trend. The mean photometric uncertainties for the target and RS are also indicated by the error bar on the second data point.

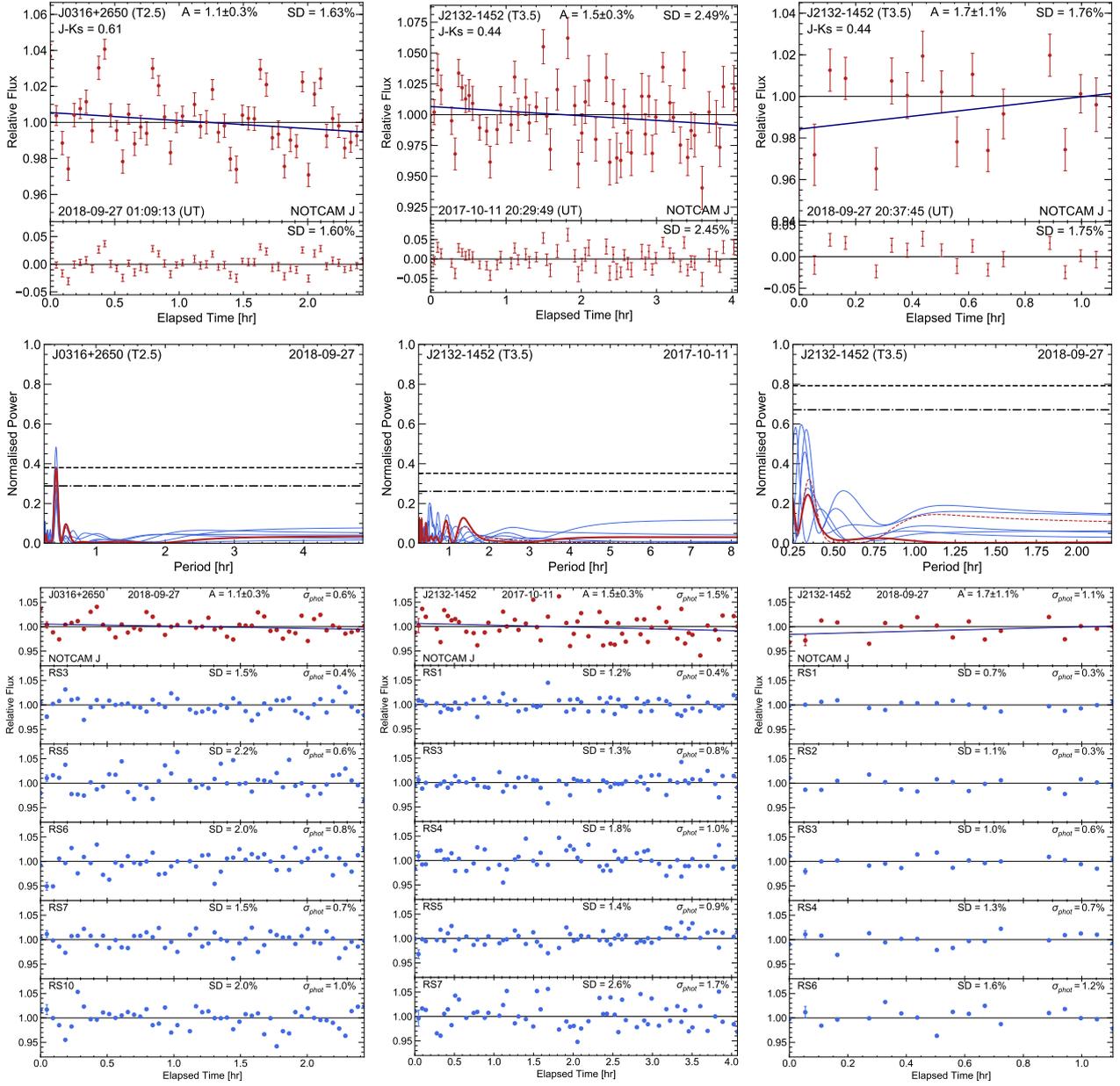

**Fig. B.2.** Same as Fig. B.1, but for the non-detection observations of J0316 in 2018 (*left set* of three figures), J2132 in 2017 (*middle set*) and 2018 (*right set*).

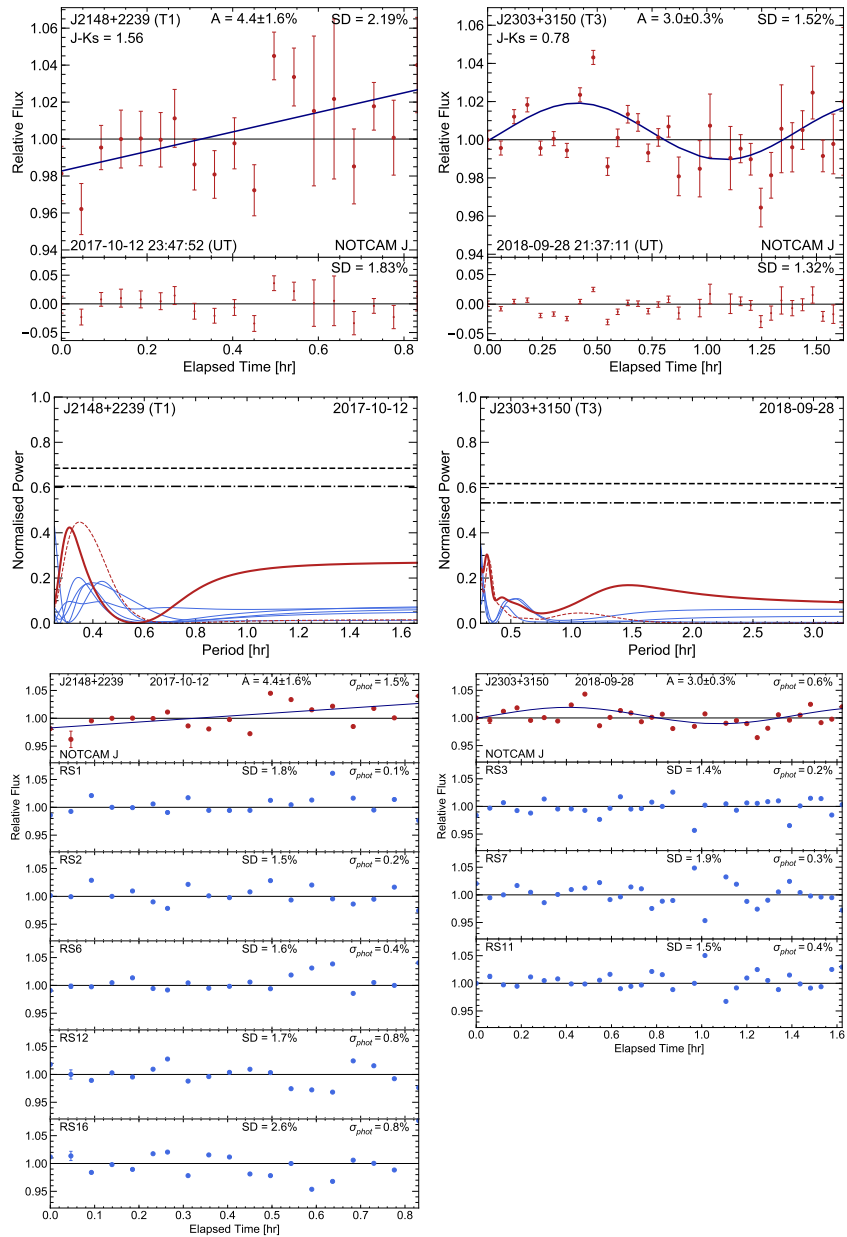

**Fig. B.3.** Same as Fig. B.1, but for the non-detection observations of J2148 in 2017 (*left set* of two figures) and J2303 in 2017 (*right set*).

## Appendix C: Finding charts for all epochs

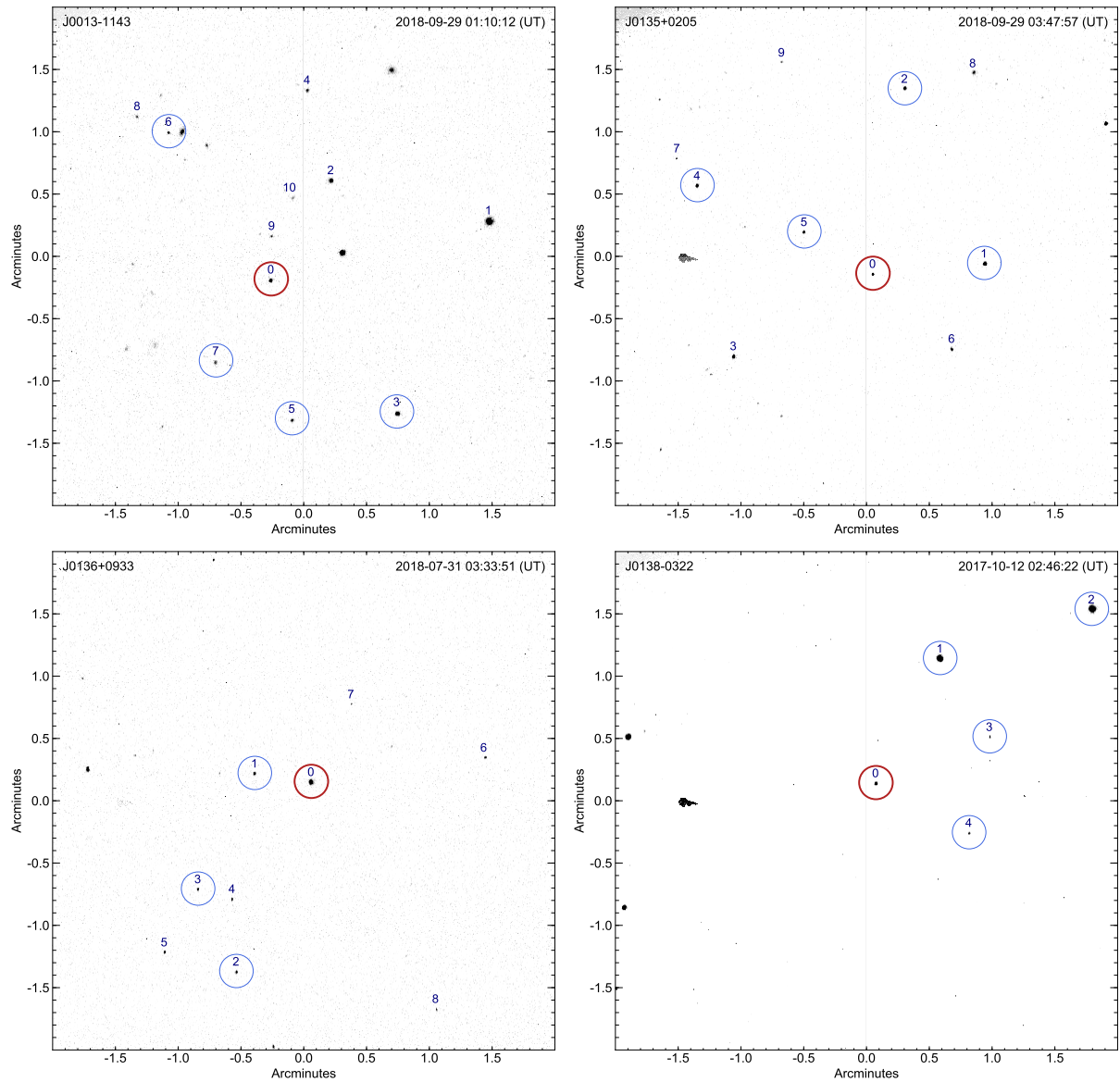

**Fig. C.1.** Finding charts for the first four epochs (J0013 in 2018 to J0138 in 2017) in Table 2. Photometry was performed on all numbered objects, starting with the target (number zero) and reference stars (numbers 1–n). The thicker red circle, radii ~30 pixels or 0.1 arcmin, indicates the target while the thinner blue circles indicate the reference stars used for the final light curves presented in Sect. 4. The faint grey column in the middle consists of dead pixels and the elongated feature on the left present in some images is a cluster of cold pixels. These bad pixel features are not visible in all finding charts due to the chosen image contrast.

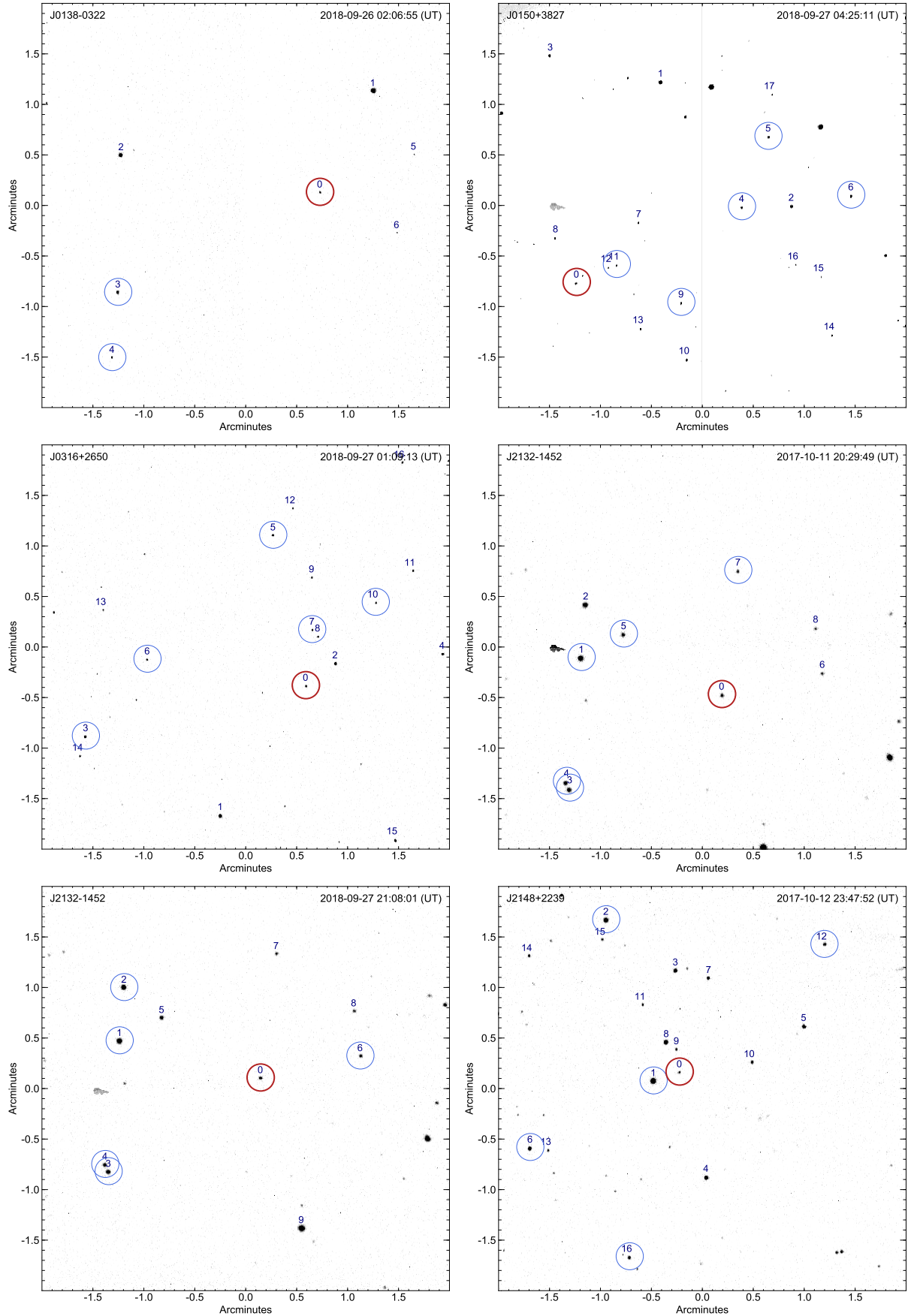

**Fig. C.2.** Same as Fig. C.1, for epochs J0138 (2018) to J2148 (2017).

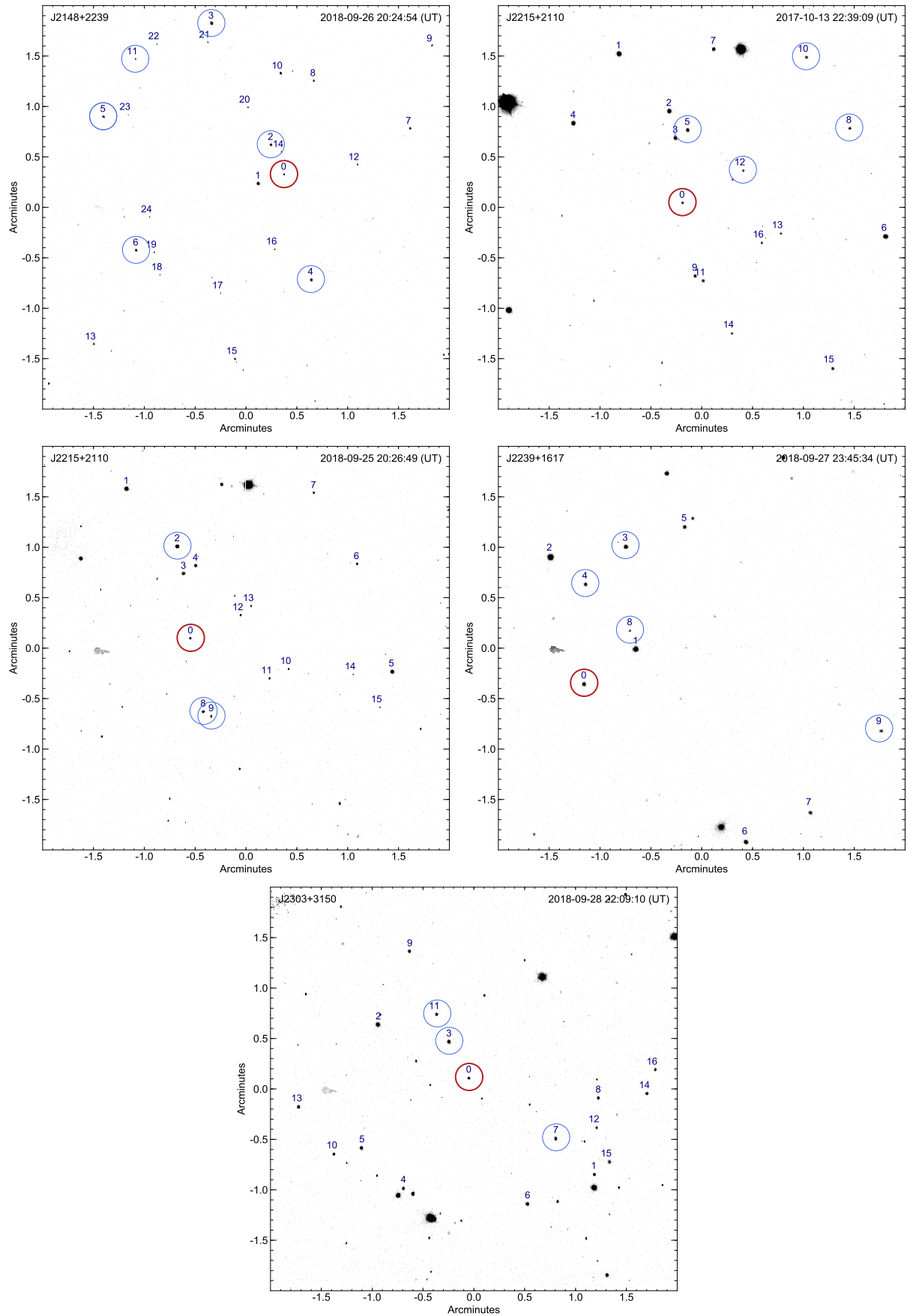

**Fig. C.3.** Same as Fig. C.1, for epochs J2148 (2018) to J2303 (2018).

## Appendix D: Additional table

**Table D.1.** Observations of known strong variables from various epochs.

| ID | SpT (NIR) | $J - K_S$ (mag) | Epoch (UT) | Amplitude [a] (%) | Period [b] (h) | $\Delta t$ (h) | Instrument [c] | Ref. |
|---|---|---|---|---|---|---|---|---|
| J0047+6803 | L7 | 2.55 | 2016-01-09 | $1.08 \pm 0.04$ | $16.3 \pm 0.2$ | 18.7 | *Spitzer*/IRAC *3.6* | 23 |
| | | | 2016-06-06 | $8.0 \pm 0.1$ [d] | $13.20 \pm 0.14$ | 8.3 | HST/WFC3 G141 *J* | 16 |
| J0501−0010 | L4 | 2.02 | 2014-11-11 | $2.0 \pm 0.1$ | 5 | 4.03 | NTT/SofI $J_S$ | 24 |
| | | | 2016-10-19 | $>1.0 \pm 0.2$ | 5 | 4.99 | NTT/SofI $J_S$ | 24 |
| J1010−0406 | L6 | 1.89 | 2012-04-06 | $3.6 \pm 0.4$ | 3 | 3.08 | NTT/SofI $J_S$ | 21 |
| J1147−2040 | L7 | 2.64 | 2017-04-17 | $1.60 \pm 0.08$ | 19.4 | 9.75 | *Spitzer*/IRAC *3.6* | 22 |
| | | | 2017-04-17 | $2.22 \pm 0.09$ | | 9.75 | *Spitzer*/IRAC *4.5* | 22 |
| 2M1207b | L5 | 1.57 | 2014-04-11 | $2.72 \pm 0.23$ | $10.7 \pm 1.2$ | 8.8 | HST/WFC3 *F125W* | 26 |
| | | | 2014-04-11 | $1.56 \pm 0.22$ | | 8.8 | HST/WFC3 *F160W* | 26 |
| J2114−2251 | L7 | 1.97 | 2014-10-09 | $10 \pm 1.3$ | $10 \pm 2$ | 5.15 | NTT/SofI $J_S$ | 24 |
| | | | 2014-11-09 | $4.8 \pm 0.7$ | 2.5 | 2.83 | NTT/SofI $J_S$ | 24 |
| | | | 2014-11-10 | $2.2 \pm 0.6$ | 4 | 3.18 | NTT/SofI $K_S$ | 24 |
| | | | 2016-08-09 | $4.8 \pm 0.2$ | $8.45 \pm 0.05$ | 9.12 | NTT/SofI $J_S$ | 24 |
| | | | 2016-08-10 | $0.96 \pm 0.08$ | | 9.65 | NTT/SofI $K_S$ | 24 |
| | | | 2016-09-08 | $3.4 \pm 0.1$ | $8.6 \pm 0.1$ | 17.2 | *Spitzer*/IRAC *4.5* | 5 |
| | | | 2016-09-08 | $5.8 \pm 0.3$ | | 7.0 | HST/WFC3 G141 *J* | 5 |
| J2244+2043 | L6 | 2.45 | 2016-07-21 | $5.5 \pm 0.6$ | 4 | 4.1 | UKIRT/WFCAM *J* | 24 |
| | | | 2016-09-15 | $0.8 \pm 0.2$ | $11 \pm 2$ | 8.8 | *Spitzer*/IRAC *3.6* | 23 |
| J0013−1143 | T3 [e] | 0.37 | 2018-09-28 | $4.6 \pm 0.2$ | 4 | 2.95 | NOT/NOTCam *J* | 1 |
| J0138−0322 | T3 | 1.06 | 2018-09-25 | $5.5 \pm 1.2$ | 4 | 4.07 | NOT/NOTCam *J* | 1 |
| J0136+0933 | T2.5 | 0.89 | 2008-09-21 | $4−6 \pm 0.5$ | $2.3895 \pm 0.0005$ | 7.4 | OMM/CPAPIR *J* | 3 |
| | | | 2008-12-13 | $5.4 \pm 0.5$ | | 3.1 | OMM/CPAPIR *J* | 3 |
| | | | 2008-12-13 | $2.6 \pm 0.5$ | | 1.9 | OMM/CPAPIR $K_S$ | 3 |
| | | | 2009-07-31 | $2.9 \pm 0.3$ | 2 | 2.32 | Du Pont/WIRC *J* | 20 |
| | | | 2011-10-06 | $3.2 \pm 0.4$ | ... | 3.15 | NTT/SofI $J_S$ | 21 |
| | | | 2011-10-10 | $4.5 \pm 0.1$ [d] | $2.39 \pm 0.07$ | 9.0 | HST/WFC3 G141 *J* | 2 |
| | | | 2013-09-28 | $1.3 \pm 0.2$ [d] | $2.414 \pm 0.078$ | 10.05 | *Spitzer*/IRAC *3.6* | 25 |
| | | | 2013-09-28 | $5.5 \pm 0.1$ [d] | | 5.3 | HST/WFC3 G141 *J* | 25 |

**Notes.** Catalogue of observations of known strong variable BDs over several epochs, listed in descending 2MASS ID (Jhh:mm±dd:mm) while sub-divided into L-, T- and Y- NIR spectral types. Catalogued BDs are also shown in the Colour ($J − K_S$) vs SpT diagram with proportional amplitudes in Fig. 7. Detailed information on colour and spectral type with references can be found in Tables 1 and 4, for new variables from this work and previously known variables respectively. [a]"Peak-to-peak" or "peak-to-trough" amplitudes. For most observations with unconstrained periods, these should be considered minimum amplitudes as the full rotation period was not covered. Amplitudes in WFC3 G141 *H* are not always explicitly listed when accompanying *J*, as typically $A_H/A_J \sim 1$, and can be found in the relevant reference if provided by the authors. [b]As in Table 3, tabulated periods should be considered to be minimum periods unless well constrained from multiple epochs. [c]Filter bandpasses for listed instruments in µm. CPAPIR: $\lambda_J = 1.25\,(1.17 − 1.33)$ and $\lambda_{K_S} = 2.15\,(2.02 − 2.30)$; GROND: $\lambda_{z'} = 0.89\,(0.81 − 1.06)$, $\lambda_J = 1.23\,(1.10 − 1.40)$, $\lambda_H = 1.63\,(1.50 − 1.80)$ and $\lambda_K = 2.16\,(1.99 − 2.35)$; IRAC: $\lambda_{3.6} = 3.18 − 3.93$ (Channel 1) and $\lambda_{4.5} = 3.99 − 5.00$ (Channel 2); MIMIR: $\lambda_J = 1.25\,(1.17 − 1.34)$; NOTCam: $\lambda_J = 1.24\,(1.17 − 1.33)$; SofI: $\lambda_{J_S} = 1.24\,(1.17 − 1.32)$ and $\lambda_{K_S} = 2.15\,(2.02 − 2.30)$; SpeX: $\lambda_J = 1.00 − 1.30$; TRAPPIST: $\lambda_{I+z} = 0.91\,(0.75 − 1.10)$; WFCAM: $\lambda_J = 1.25\,(1.17 − 1.33)$; WFC3 G141: $\lambda_J = 1.21 − 1.32$ (narrow *J*) and $\lambda_H = 1.54 − 1.60$ (narrow *H*); WIRC: $\lambda_J = 1.24\,(1.14 − 1.33)$. [d]$1\sigma$ amplitude uncertainty estimated from reference, or for observations done using HST or *Spitzer* the uncertainty ($\pm0.1\%$ and $\pm0.2\%$ respectively) is taken from an average of tabulated uncertainties. [e,f]Possible binary or resolved binary respectively, as tabulated in Tables 1 and 4.

**Table D.5.** continued.

| ID | SpT (NIR) | $J - K_S$ (mag) | Epoch (UT) | Amplitude [a] (%) | Period [b] (h) | $\Delta t$ (h) | Instrument [c] | Ref. |
|---|---|---|---|---|---|---|---|---|
| | | | 2015-11-10 | $4.0 \pm 0.3$ [d] | $2.406 \pm 0.008$ | 8.82 | Perkins/MIMIR $J$ | 10 |
| | | | 2015-11-12 | $<1.0 \pm 0.3$ [d] | | 9.47 | Perkins/MIMIR $J$ | 10 |
| | | | 2015-11-13 | $3.0 \pm 0.3$ [d] | | 8.82 | Perkins/MIMIR $J$ | 10 |
| | | | 2018-07-31 | $4.4 \pm 0.2$ | $2.14 \pm 0.05$ | 2.72 | NOT/NOTCam $J$ | 1 |
| J0447−1216 | T2 | 0.93 | 2016-10-18 | $4.5 \pm 0.6$ | 3 | 3.31 | NTT/SofI $J_S$ | 24 |
| | | | 2017-12-08 | ... | ... | 4.29 | UKIRT/WFCAM $J$ | 24 |
| J0758+3247 | T2 [e] | 0.74 | 2009-12-26 | $4.8 \pm 0.2$ | $4.9 \pm 0.2$ | 2.9 | Du Pont/WIRC $J$ | 20 |
| | | | 2010-03-10 | $<0.9 \pm 0.8$ | ... | 5.3 | OMM/CPAPIR $J$ | 14 |
| J1049−5319 B | T0.5 | 1.49 | 2013-03-20 | $11 \pm 1$ [d] | $4.87 \pm 0.01$ | 7.9 | TRAPPIST $I + z$ | 13 |
| | | | 2013-03-21 | $6 \pm 1$ [d] | ... | 8.2 | TRAPPIST $I + z$ | 13 |
| | | | 2013-04-22 | $7 \pm 0.5$ [d] | ... | 3.2 | MPG/GROND $z'$ | 4 |
| | | | 2013-04-22 | $<3$ | ... | 3.2 | MPG/GROND $J$ | 4 |
| | | | 2013-04-22 | $13 \pm 2$ [d] | ... | 3.2 | MPG/GROND $H$ | 4 |
| | | | 2013-04-22 | $10 \pm 2$ [d] | ... | 3.2 | MPG/GROND $K$ | 4 |
| | | | 2013-04-26 | $5 \pm 1$ [d] | $5.05 \pm 0.10$ | 7.5 | TRAPPIST $I + z$ | 8 |
| | | | 2013-04-30 | $\sim 7.5$ | ... | 0.75 | IRTF/SpeX $J$ | 8 |
| | | | 2013-11-08 | $\sim 10$ | ... | 6.6 | HST/WFC3 G141 $J$ | 7 |
| | | | 2013-11-08 | $\sim 9$ | ... | 6.6 | HST/WFC3 G141 $H$ | 7 |
| J1052+4422 | T0.5 [f] | 1.39 | 2010-03-12 | $2.2 \pm 0.8$ | 2.5 | 6.0 | OMM/CPAPIR $J$ | 14 |
| | | | 2010-03-15 | $3.6 \pm 0.8$ | 3 | 5.0 | OMM/CPAPIR $J$ | 14 |
| Ross 458C | T8 | -0.21 | 2012-09-04 | $<1.37$ | ... | 13.8 | $Spitzer$/IRAC 3.6 | 17 |
| | | | 2012-09-04 | $<0.72$ | ... | 6.9 | $Spitzer$/IRAC 4.5 | 17 |
| | | | 2018-01-06 | $2.62 \pm 0.02$ | $6.75 \pm 1.58$ | 10 | HST/WFC3 G141 $J$ | 17 |
| J1324+6358 | T2 [e] | 1.54 | 2012-04-22 | $3.05 \pm 0.15$ | $13 \pm 1$ | 13.8 | $Spitzer$/IRAC 3.6 | 18 |
| | | | 2012-04-22 | $3.0 \pm 0.3$ | | 6.9 | $Spitzer$/IRAC 4.5 | 18 |
| | | | 2013-04-20 | $5 - 8 \pm 0.2$ [d] | ... | 54.4 | $Spitzer$/IRAC 3.6 | 25 |
| | | | 2014-04-19 | $4.7 \pm 0.2$ [d] | ... | 12.3 | $Spitzer$/IRAC 4.5 | 25 |
| | | | 2014-04-20 | $7.3 \pm 0.2$ [d] | ... | 14.5 | $Spitzer$/IRAC 3.6 | 25 |
| | | | 2014-05-10 | $2 - 12 \pm 0.2$ [d] | ... | 55.3 | $Spitzer$/IRAC 3.6 | 25 |
| J1516+3053 | T0.5 [e] | 1.61 | 2012-05-06 | $2.4 \pm 0.2$ | 6.7 | 13.4 | $Spitzer$/IRAC 3.6 | 18 |
| | | | 2012-05-06 | $3.1 \pm 0.2$ | | 6.9 | $Spitzer$/IRAC 4.5 | 18 |
| J1629+0335 | T2 | 1.25 | 2009-07-29 | $4.3 \pm 2.4$ | $6.9 \pm 2.4$ | 4.03 | Du Pont/WIRC $J$ | 20 |
| J2139+0220 | T1.5 [e] | 1.13 | 2009-08-02 | $9 \pm 1.0$ | ... | 2.5 | Du Pont/WIRC $J$ | 20 |
| | | | 2009-09-23 | $26 \pm 1.0$ | $7.721 \pm 0.005$ | 6.0 | Du Pont/WIRC $J$ | 19 |
| | | | 2009-09-30 | $10.9 \pm 1.1$ | | 4.8 | Du Pont/WIRC $J$ | 18 |
| | | | 2009-09-30 | $9.2 \pm 1.5$ | | 4.8 | Du Pont/WIRC $H$ | 18 |
| | | | 2009-09-30 | $6.4 \pm 2.0$ | | 4.8 | Du Pont/WIRC $K_S$ | 18 |
| | | | 2010-10-21 | $27 \pm 0.10$ | $7.8 \pm 0.1$ | 9.0 | HST/WFC3 G141 $J$ | 2 |
| | | | 2011-10-04 | $5.9 \pm 0.4$ | ... | 3.3 | NTT/SofI $J_S$ | 21 |
| | | | 2013-02-03 | $8 \pm 0.2$ [d] | $7.614 \pm 0.178$ | 31.5 | $Spitzer$/IRAC 3.6 | 25 |
| | | | 2014-01-23 | $11 \pm 0.2$ [d] | | 31.5 | $Spitzer$/IRAC 3.6 | 25 |
| | | | 2014-02-09 | $4 \pm 0.2$ [d] | | 31.5 | $Spitzer$/IRAC 3.6 | 25 |
| HN Peg B | T2.5 | 0.74 | 2012-01-29 | $0.77 \pm 0.15$ | $18 \pm 4$ | 13.8 | $Spitzer$/IRAC 3.6 | 18 |
| | | | 2012-01-29 | $1.11 \pm 0.5$ | | 6.9 | $Spitzer$/IRAC 4.5 | 18 |
| | | | 2016-07-11 | ... | ... | 5.0 | UKIRT/WFCAM $J$ | 24 |
| | | | 2016-07-13 | ... | ... | 5.0 | UKIRT/WFCAM $J$ | 24 |
| | | | 2017-05-16 | $2.56 \pm 0.03$ | $15.4 \pm 0.5$ | 8.5 | HST/WFC3 G141 $J$ | 27 |
| J2215+2110 | T1 [e] | 1.18 | 2017-10-13 | $10.7 \pm 0.4$ | $3.0 \pm 0.2$ | 2.4 | NOT/NOTCam $J$ | 1 |
| | | | 2018-09-25 | $2.6 \pm 0.2$ | $5.2 \pm 0.5$ | 5.1 | NOT/NOTCam $J$ | 1 |
| J2228−4310 | T6 | 0.36 | 2008-10-03 | $1.54 \pm 0.14$ | $1.43 \pm 0.16$ | 7.5 | NTT/SofI $J_S$ | 9 |
| | | | 2009-08-01 | $1.6 \pm 0.3$ | $1.42 \pm 0.03$ | 5.73 | Du Pont/WIRC $J$ | 20 |

**Table D.5.** continued.

| ID | SpT (NIR) | $J - K_S$ (mag) | Epoch (UT) | Amplitude [a] (%) | Period [b] (h) | $\Delta t$ (h) | Instrument [c] | Ref. |
|---|---|---|---|---|---|---|---|---|
| | | | 2011-07-07 | $1.85 \pm 0.07$ | $1.405 \pm 0.005$ | 9.6 | HST/WFC3 G141 *J* | 6 |
| | | | 2011-07-07 | $2.74 \pm 0.11$ | | 9.6 | HST/WFC3 G141 *H* | 6 |
| | | | 2011-07-07 | $1.47 \pm 0.14$ | $1.38 \pm 0.03$ | 5.5 | *Spitzer*/IRAC *4.5* | 6 |
| | | | 2012-08-14 | $4.6 \pm 0.2$ | $1.41 \pm 0.01$ | 13.8 | *Spitzer*/IRAC *3.6* | 18 |
| | | | 2012-08-14 | $1.51 \pm 0.15$ | | 6.9 | *Spitzer*/IRAC *4.5* | 18 |
| | | | 2013-07-20 | $4.7 \pm 0.2$ [d] | $1.369 \pm 0.032$ | 7.95 | *Spitzer*/IRAC *3.6* | 25 |
| | | | 2013-07-20 | $2.00 \pm 0.2$ [d] | | 2.1 | *Spitzer*/IRAC *4.5* | 25 |
| | | | 2013-07-20 | $2.1 \pm 0.1$ [d] | | 6.5 | HST/WFC3 G141 *J* | 25 |
| | | | 2013-07-27 | $4.2 \pm 0.2$ [d] | | 3.6 | *Spitzer*/IRAC *3.6* | 25 |
| | | | 2013-07-27 | $1.6 \pm 0.1$ [d] | | 5.2 | HST/WFC3 G141 *J* | 25 |
| J2239+1617 | T3 | 1.19 | 2018-09-27 | $5.8 \pm 0.4$ | 4 | 2.64 | NOT/NOTCam *J* | 1 |
| J0855−0714 | >Y2 | … | 2015-03-09 | $4-5 \pm 0.2$ [d] | 5 | 11.5 | *Spitzer*/IRAC *4.5* | 12 |
| | | | 2015-03-10 | $4-5 \pm 0.4$ [d] | 5 | 11.5 | *Spitzer*/IRAC *3.6* | 12 |
| J1405+5534 | Y0.5 | −0.55 | 2013-08-17 | $7.08 \pm 0.09$ | $8.54 \pm 0.08$ | 12.2 | *Spitzer*/IRAC *4.5* | 11 |
| | | | 2013-08-17 | $7.2 \pm 0.4$ | $8.2 \pm 0.3$ | 12 | *Spitzer*/IRAC *3.6* | 11 |
| J1738+2732 | Y0 | −1.00 | 2013-06-30 | $3.0 \pm 0.2$ [d] | $6.0 \pm 0.1$ | 12 | *Spitzer*/IRAC *4.5* | 15 |